\documentclass[aps,prx,superscriptaddress,twocolumn,floatfix,citeautoscript,longbibliography,hyperlinks]{revtex4-1}

\usepackage{graphicx}
\usepackage{dcolumn}
\usepackage[dvipsnames]{xcolor}
\usepackage[colorlinks=true,urlcolor=blue,citecolor=blue, linkcolor = blue]{hyperref}
\usepackage{amssymb,amsmath,bm,bbm,mleftright,mathtools}
\DeclareMathAlphabet\mathbfcal{OMS}{cmsy}{b}{n}
\newcommand{\bra}[1]{\left\langle #1\right|}
\newcommand{\ket}[1]{\left|#1\right\rangle}

\renewcommand{\vec}{\mathbf}
\makeatletter
\newsavebox{\@brx}
\newcommand{\llangle}[1][]{\savebox{\@brx}{\(\m@th{#1\langle}\)}%
  \mathopen{\copy\@brx\kern-0.5\wd\@brx\usebox{\@brx}}}
\newcommand{\rrangle}[1][]{\savebox{\@brx}{\(\m@th{#1\rangle}\)}%
  \mathclose{\copy\@brx\kern-0.5\wd\@brx\usebox{\@brx}}}
\makeatother

\begin{document}

\title{Boiling Quantum Vacuum: Thermal Subsystems from Ground-State Entanglement}

\author{ Ali G. Moghaddam}
\affiliation{Computational Physics Laboratory, Physics Unit, Faculty of Engineering and
Natural Sciences, Tampere University, FI-33014 Tampere, Finland}
\affiliation{Helsinki Institute of Physics, FI-00014 University of Helsinki, Finland}
\author{ Kim P\"oyh\"onen}
\affiliation{Computational Physics Laboratory, Physics Unit, Faculty of Engineering and
Natural Sciences, Tampere University, FI-33014 Tampere, Finland}
\affiliation{Helsinki Institute of Physics, FI-00014 University of Helsinki, Finland}
\author{Teemu Ojanen} \email{Email: teemu.ojanen@tuni.fi}
\affiliation{Computational Physics Laboratory, Physics Unit, Faculty of Engineering and
Natural Sciences, Tampere University, FI-33014 Tampere, Finland}
\affiliation{Helsinki Institute of Physics, FI-00014 University of Helsinki, Finland}

\begin{abstract}
In certain special circumstances, such as in the vicinity of a black hole or in a uniformly accelerating frame, vacuum fluctuations appear to give rise to a finite-temperature environment. This effect, currently without experimental confirmation, can be interpreted as a manifestation of quantum entanglement after tracing out vacuum modes in an unobserved region. In this work, we identify a class of experimentally accessible quantum systems where thermal density matrices emerge from vacuum entanglement. We show that reduced density matrices of lower-dimensional subsystems embedded in $D$-dimensional gapped Dirac fermion vacuum, either on a lattice or continuum, have a thermal form with respect to a lower-dimensional Dirac Hamiltonian. Strikingly, we show that vacuum entanglement can even conspire to make a subsystem of a gapped system at zero temperature appear as a hot gapless system. We propose concrete experiments in cold atom quantum simulators to observe the vacuum entanglement induced thermal states.            

\end{abstract}

\maketitle

\section{Introduction}

Thermalization is one of the most widespread and fundamental phenomena and plays a central role in virtually all branches of physics. In standard textbook statistical physics, a thermal state arises as a maximum entropy state that satisfies appropriate external constraints \cite{kardar2007book}. More recently, the notion of the eigenstate thermalization hypothesis has identified temperature as a generic emergent phenomenon in closed quantum systems \cite{Deutsch1991,Srednicki1994,Srednicki2012}. According to the hypothesis, 
the reduced density matrix of a subsystem 
of a thermodynamically large, interacting many-body system 
is asymptotically equal to the thermal reduced density matrix
when the subsystem is sufficiently small compared to the total system \cite{rigol2008thermalization, Garrison2018,Polkovnikov2016ETH}.
This hypothesis in its strong form, where all eigenstates become thermalized,
has been verified for nonintegrable systems \cite{Deutsch_2018}.
A weaker version, such that an exponentially small number of nonthermal states 
can exist, has been observed in certain integrable models as well \cite{Lauchli2010}.

While the eigenstate thermalization hypothesis only accounts for a non-zero temperature in highly excited systems, there are famous examples of how vacuum fluctuations may give rise to a finite temperature environment. The Hawking effect, which attributes a finite temperature to black holes, is deeply connected to the entanglement of vacuum modes \cite{Sorkin1986,Srednicki1993,Frolov1998}. In the same vein, the Unruh effect gives rise to a finite temperature for accelerated observers moving in the relativistic vacuum. In both cases, the apparent unitarity-violating emergence of a thermal state could be attributed to entanglement with an unobservable region beyond the event horizon or the Rindler wedge. 
\begin{figure}[t]
    \centering
    \includegraphics[width=0.99\columnwidth]{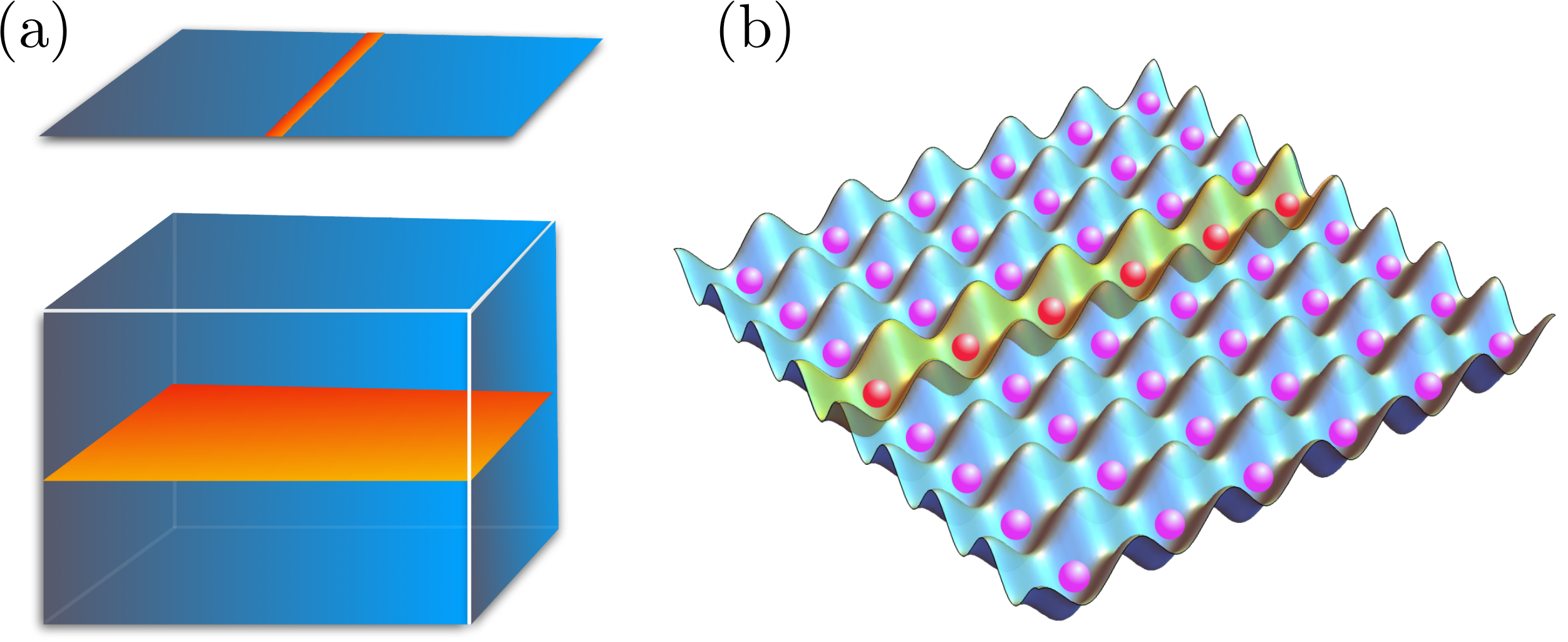}
    \caption{Entanglement-induced thermal subsystems embedded in $D$-dimensional Dirac fermion system at zero temperature. (a): Examples of lower-dimensional thermal subsystems embedded in 2D and 3D parent states. When the full system is in the ground state of the total system Hamiltonian $\mathcal{H}_D$, the reduced density matrix of the subsystem has a thermal form $\rho=e^{-\beta\mathcal{H}_{D-1}}/\mathcal{Z}$. (b): Thermal state emerging from vacuum entanglement could be observed in cold atom quantum simulators by probing particle fluctuations in the one-dimensional subsystem.   }
    \label{fig:1}
\end{figure}
It has been recognized that this picture is valid in a much broader sense, promoting entanglement as the key unifying concept in analyzing diverse phenomena from black hole physics to condensed matter systems \cite{solodukhin2011review,ryu2006,Casini2009,osterloh2002scaling,Kitaev2003,CalabreseCardy2004,terhal2003entanglement,vedral2008rmp,Horodecki,Eisert2010}.
For example, the emergence of effective temperature from ground-state entanglement has lately been identified in systems obeying the entanglement \emph{area law}, where the entanglement entropy scales as the subsystem boundary \cite{hastings2007area,Plenio2005,Eisert2010}. The area law is known to give rise to subsystem density matrices which are characterized by a spatially-varying effective temperature that decreases rapidly away from the boundaries \cite{BW1975,Swingle2016,dalmonte2018quantum,dalmoste2022,vaezi2021}. Unfortunately, the strongly inhomogeneous entanglement temperature profile is mostly of theoretical interest since its experimental verification poses so far unresolved practical and conceptual issues. However, a direct experimental observation of a thermal state emerging from vacuum entanglement would be an outstanding achievement with deep implications for multiple branches of physics.

In this work, we identify a large class of systems, illustrated in Fig.~\ref{fig:1}(a), where vacuum entanglement induces a uniform temperature and where the phenomenon becomes experimentally accessible. Specifically, we show that the lower-dimensional subsystems embedded in a $D$-dimensional gapped Dirac fermion vacuum have thermal density matrices. This property holds for continuum models as well as for lattice systems. The thermal Hamiltonian of a subsystem has a simple relation to the Hamiltonian of the whole system, while the effective temperature is determined by the bandwidth in the traced-out directions. For lattice systems, the effective temperature acquires momentum dependence; however, typically the density matrix is excellently reproduced by a constant-temperature approximation. We explain how the notion of lower-dimensional thermal subsystems is closely connected to the table of topological insulators in different dimensions. As a striking consequence of our results, we show that the vacuum entanglement can conspire to make lower-dimensional subsystems of a zero-temperature gapped state appear as hot gapless systems. Finally, we explain how the thermal nature of the subsystems manifests through fluctuations in observables and propose a concrete setup, illustrated in Fig. \ref{fig:1}(b), where our predictions can be verified in cold atom quantum simulators. Specifically, we show that the particle number fluctuations in a one-dimensional chain embedded in a two-dimensional array match those of a genuinely one-dimensional Dirac system at finite temperature, providing a smoking gun signature of the vacuum entanglement-induced thermal state.

\section{Thermal entanglement spectra in $D$-dimensional gapped Fermi systems} 

In this section, we study lower-dimensional subsystems embedded in the ground state of a gapped $D$-dimensional Dirac fermion system with the Hamiltonian 
\begin{align}\label{eq:dirac}
H_D({\bf{k}}) = \sum_{\mu}d_{D\mu}({\bf k}) \Gamma^\mu\equiv\vec{d}_D({\bf k})\cdot\boldsymbol{\Gamma},    
\end{align}
where $\Gamma^\mu$ are $2^n$-dimensional (with $n\in {\mathbbm N}$)  Clifford matrices $\{\Gamma^\mu,\Gamma^\nu\}=2{\mathbbm 1}\delta^{\mu\nu}$ and $\vec{d}_D$ satisfies $\vec{d}_D\cdot\vec{d}_D>0$ for all $D$-dimensional (quasi)momenta $\vec{k}\in \mathbb{R}^D$. In particular, we show that the reduced density matrix of a $D_s$-dimensional translation-invariant subsystem $(D_s<D)$ can be exactly written in a form as 
\begin{align}\label{eq:reduced}
\rho_{D_s}=\frac{e^{-\sum_{k_s}\beta(\vec{k}_s) \hat{d}^\dagger_{\vec{k}_s}H_{D_s}({\bf{k}_s})\hat{d}_{\vec{k}_s}}}{\mathcal{Z}},
\end{align}
where the effective subsystem Hamiltonian (ESH) $H_{D_s}({\bf{k}_s})=\vec{d}_{D_s}(\vec{k}_s)\cdot\boldsymbol{\Gamma}$ has a Dirac form with a lower-dimensional momentum $\vec{k}_s\in \mathbb{R}^{D_s}$ and $\hat{d}_{\vec{k}_s}$ are fermion annihilation operators. 
The reduced density matrix in Eq. \eqref{eq:reduced} 
in its general form corresponds to a \emph{generalized Gibbs ensemble}
which reduces to an exactly thermal density matrix
for a constant $\beta$ \cite{Olshanii2007}.
We obtain an analytical expression for the effective translation-invariant inverse temperature $\beta(\vec{k}_s)$ and demonstrate with examples how expression \eqref{eq:reduced} typically holds to remarkable accuracy when $\beta(\vec{k}_s)$ is approximated by a constant. Despite the system as a whole being in the quantum ground state, from the point of view of observables, the subsystems behave as $D_s$-dimensional systems at finite temperature. We note that the ESH should not be confused with the commonly-studied entanglement Hamiltonian $H_E$, defined by $\rho_{D_s}=e^{-H_E}/{\cal Z}$. In contrast to the ESH, the entanglement Hamiltonian does not provide a natural notion of temperature, and it does not reduce to the subsystem Hamiltonian even when all couplings between the reduced subsystem and the rest vanish.  

\subsection{Entanglement-Temperature mapping}

Here we derive the entanglement-temperature mapping in Eq.~\eqref{eq:reduced}. For a free fermion system in a Gaussian state, including (but not limited to) the ground state and a finite-temperature state, the reduced density matrix of an arbitrary subsystem also corresponds to a Gaussian state \cite{Peschel_2003,Peschel_2009}. Consequently, due to Wick's theorem, the entanglement spectrum of a subsystem is completely encoded in the correlation matrix with real-space components defined as
$\mathcal{C}_{{\bf x},{\bf x}^\prime}^{\alpha\alpha^\prime}=\langle\hat{c}^\dagger_{{\bf x}\alpha}\hat{c}_{{\bf x}^\prime\alpha^\prime}\rangle^{\ast}$ given in terms of fermion operators $\hat{c}_{{\bf x}\alpha}$ for a particle with orbital index $\alpha$ and at position $x$ in the subsystem. 
If two systems have the same correlation matrices, they necessary have coinciding reduced density matrices. Here, by matching the correlation matrices, we map the reduced density matrix of $D_s$-dimensional subsystems to thermal $D_s$-dimensional systems given by Eq.~\eqref{eq:reduced}.  
In translationally invariant systems, by expanding particle creation operators
in the basis of Bloch eigenstates $\psi_{\nu{\bf k}}$ as $\hat{c}^\dag_{\vec{k}\alpha}=\sum_{\nu}  \langle\alpha|\psi_{\nu{\bf k}}\rangle  \hat{d}^\dagger_{\nu{\bf k}} $ we find the correlation matrix elements in $k$-space as
\begin{equation} \nonumber
    \langle\hat{c}^\dagger_{{\bf k}\alpha}\hat{c}_{{\bf k}\alpha^\prime}\rangle^{\ast} = \sum_{\nu}\langle\alpha|\psi_{\nu{\bf k}}\rangle
    \langle\psi_{\nu{\bf k}}|\alpha^\prime\rangle \langle\hat{d}^\dagger_{\nu{\bf k}}\hat{d}_{\nu{\bf k}}\rangle ,
\end{equation}
where the expectation value on the right hand side gives the Fermi-Dirac distribution $n_{F}(E_{\nu{\bf k}})$. In the following, we assume that the parent $D$-dimensional system is at zero temperature so $n_{F}(E_{\nu{\bf k}})$ is 1 for filled bands and 0 for others. It is now straightforward to show that by restricting spatial indices ${\bf x},{\bf x}'$ to a $D_s$-dimensional subsystem with periodic boundary conditions, the correlation matrix becomes
\begin{align}\label{reducedC}
	\mathcal{C}_{{\bf x},{\bf x}'}^{\alpha\alpha'}
&
=
\frac{1}{L^{D}}\sum_{\vec{k}}
e^{-i\vec{k}\cdot({\bf x}-{\bf x}')}
\langle\hat{c}^\dagger_{{\bf k}\alpha}\hat{c}_{{\bf k}\alpha^\prime}\rangle^{\ast}
\nonumber\\	
&=
\frac{1}{L^{D_s}}\sum_{\vec{k}_s}e^{-i\vec{k}_s\cdot({\bf x}-{\bf x}')}\bra{\alpha}\hat{\mathbfcal C}^{\rm sub}({\bf k}_{s})\ket{\alpha'},
\end{align}
where $L$ is the linear extent of the system in all $D$ dimensions and
\begin{equation}\label{eq:subsys-general-corr}
	\hat{\mathbfcal C}^{\rm sub}({\bf k}_{s})= \frac{1}{L^{D-D_s}}
	\sum_{{\rm filled}~\nu,{\bf k}_{\perp}} |\psi_{\nu{\bf k}}\rangle \langle \psi_{\nu{\bf k}}|.
\end{equation}
This defines the Fourier transform of subsystem correlation matrix which has been obtained simply by plugging the above expression for $\langle\hat{c}^\dagger_{{\bf k}\alpha}\hat{c}_{{\bf k}\alpha^\prime}\rangle^{\ast}$.
The full $D$-dimensional momentum $\vec{k}=(\vec{k}_s,\vec{k}_{\perp})$ is decomposed as the reduced subsystem momentum ${\bf k}_s$ with $D_s$ components, and the momentum  perpendicular to the subsystem ${\bf k}_\perp$ with $D-D_s$ components. 
We note that, since the Hamiltonian
is expressed in terms of anticommuting gamma matrices, the number of different orbitals
(bands) $\alpha$ are also limited to $2^n$
and 
the correlation matrix $\hat{\mathbfcal C}$
must be $2^n\times 2^n$, accordingly.
The entanglement spectrum and the reduced density matrix are now fully determined by the correlation matrix \eqref{reducedC}.

The correlation matrix of a genuinely $D_s$-dimensional system at a finite temperature is also given by expression \eqref{reducedC} but now with operator $\hat{\mathbfcal C}^{\rm sub}({\bf k}_{s})$ substituted by
\begin{equation}\label{eq:thermal-general-corr}
  \hat{\mathbfcal C}^{\rm th}({\bf k}_{s})= \sum_{\nu} |\phi_{\nu{\bf k}_s}\rangle \langle \phi_{\nu{\bf k}_s}| n_F(\omega_{{\nu\bf k}_s}).
\end{equation}
where $\ket{\phi_{\nu{\bf k}_s}}$ and $\omega_{{\nu\bf k}}$ are  eigenstates and energies of a $D_s$-dimensional Hamiltonian. The necessary and sufficient condition for the thermal mapping of the reduced density matrix of $D_s$-dimensional subsystems is that expressions \eqref{eq:subsys-general-corr} and \eqref{eq:thermal-general-corr} must match for some $D_s$-dimensional Hamiltonian $H_{D_s}$. Thus, the emergence of an effective temperature in the subsystem reduced density matrix arises from the momentum average of $D$-dimensional band projectors over the $D-D_{s}$ unobserved dimensions. Eqs.~\eqref{reducedC}-\eqref{eq:thermal-general-corr} are valid for all free fermions systems. 

We now show how the generalized Dirac systems \eqref{eq:dirac} provide a natural example of entanglement-temperature correspondence \eqref{eq:reduced}. The spectrum of the $D$-dimensional parent Hamiltonian \eqref{eq:dirac} is given by $\varepsilon_{\bf k}=\pm|{\bf d}_D({\bf k})|$ and the projection to the filled negative-energy bands is obtained by
\begin{equation}\nonumber
    \sum_{{\rm filled}~\nu}|\psi_{\nu{\bf k}}\rangle \langle \psi_{\nu{\bf k}}|=\frac{1}{2}\big( {\mathbbm 1} - \frac{{\bf d}_D}{|{\bf d}_D|}\cdot {\bm \Gamma} \big).
\end{equation}
Hence, we find
\begin{equation}\label{eq:sub-corr-Dirac}
   \hat{\mathbfcal C}^{\rm sub}({\bf k}_{s}) = 
   \frac{1}{2}\big( {\mathbbm 1} - \big\langle \frac{{\bf d}_D}{|{\bf d}_D|} \big\rangle_{\perp}\cdot {\bm \Gamma} \big),
\end{equation}
with
\begin{equation}\nonumber
 \langle\cdots\rangle_{\perp}=L^{-(D-D_s)}\sum_{{\vec{k}}_\perp}\cdots
\end{equation}
denoting the momentum average over the traced over dimensions. Defining a new quantity 
\begin{equation}\label{eq:ds}
{\bf d}_{D_s}(\vec{k}_s)	
	=\frac{1}{{\cal F}_{D_s}({\bf k}_{s})} \big\langle \frac{{\bf d}_D}{|{\bf d}_D|} \big\rangle_{\perp},
\end{equation}
where 
\begin{align}\label{eq:F}
	{\cal F}_{D_s}({\bf k}_{s})&=\langle \frac{1}{|{\bf d}_D|} \rangle_{\perp}, 
\end{align}
the correlation matrix \eqref{eq:sub-corr-Dirac} for the reduced system becomes 
\begin{equation}\label{eq:sub-corr-Dirac2}
   \hat{\mathbfcal C}^{\rm sub}({\bf k}_{s}) = 
   \frac{1}{2}\big( {\mathbbm 1} - {\cal F}_{D_s}{\bf d}_{D_s}\cdot {\bm \Gamma} \big).
\end{equation}
This is immediately similar to the thermal correlation matrix 
of a genuinely $D_s$-dimensional system with a Dirac Hamiltonian $H_{D_s}=\vec{d}_{D_s}({\bf k}_s)\cdot\boldsymbol{\Gamma}$. Using Eq.~\eqref{eq:thermal-general-corr}, the thermal correlation matrix for such a system reads
\begin{equation}\nonumber
   \hat{\mathbfcal C}^{\rm th}({\bf k}_{s}) =
   \frac{1}{2}\sum_{\eta=\pm1}\Big( {\mathbbm 1} + \eta 
   \frac{{\bf d}_{D_s}}{|{\bf d}_{D_s}|}\cdot {\bm \Gamma}
   \Big)n_F(\eta|{\bf d}_{{\bf k}_s}|),
\end{equation}
which can be matched with \eqref{eq:sub-corr-Dirac2} by requiring
\begin{equation}
{\cal F}_{D_s}({\bf k}_{s})=\frac{n_F(-|{\bf d}_{D_s}|)-n_F(|{\bf d}_{D_s}|)}{|{\bf d}_{D_s}|}.\nonumber
\end{equation}
From this equation we can solve the effective entanglement temperature as
\begin{equation}\label{eq:T_eff}
    T(\vec{k}_s)=\beta^{-1}(\vec{k}_s)= \frac{|{\bf d}_{D_s}|}{2\:{\rm arctanh}(|{\bf d}_{D_s}|{\cal F}_{D_s})},
\end{equation}

The ESH \eqref{eq:ds} and temperature \eqref{eq:T_eff} fix the entanglement-temperature mapping in Eq.~\eqref{eq:reduced}, proving that the ground state entanglement in lower-dimensional subsystems give rise to a thermal density matrix. This density matrix is characterized by a translation-invariant temperature and the ESH $H_{D_s}$ which is obtained by averaging the parent Hamiltonian $H_D$ over the unobserved directions. The entanglement temperature \eqref{eq:T_eff} is of the order of the bandwidth (or the hopping amplitude) in the traced-out dimensions and, as such, very high for isotropic models. In strong contrast to generic area-law subsystems which exhibit strongly inhomogeneous spatial temperature profile \cite{dalmonte2018quantum,dalmoste2022}, the entanglement entropy here scales as the subsystem volume, and the effective temperature for lower-dimensional systems can be typically regarded as a constant as seen below. The analogy to a true thermal equilibrium state with uniform temperature makes the phenomenon feasible to experimental studies.

\subsection{Example I: 1D thermal subsystems in a Chern insulator}\label{subsecChern}
To make the general entanglement-temperature mapping more concrete, we now illustrate it by examples. First we study a 2D Chern insulator model and show that its 1D subsystems corresponds to thermal 1D systems. In particular, we consider the Qi-Wu-Zhang (QWZ) model defined by $H_{2D}= \vec d_{2D}(\vec{k})\cdot{\bm\sigma}$
with $   \vec d_{2D}(\vec{k})=(t_x\sin k_x, t_y\sin k_y, m-t_x\cos k_x -t_y\cos k_y)$.
For the sake of compactness, in the following we set $t_x=1$, which is equivalent to measuring all other energy scales with respect to that quantity. The correlation matrix given by Eq. \eqref{eq:subsys-general-corr}, for a 1D subsystem in $x$ direction for the 2D model with the valence band filled reads
\begin{align}\nonumber
      {\mathbfcal C}(k_x)  &=
      \frac{1}{L}\sum_{k_y} \ket{\psi_{{\bf k},-}} \bra{\psi_{{\bf k},-}}
      \nonumber\\
      &=
       \frac{1}{2}\Big[ {\mathbbm 1} - \frac{1}{L} \sum_{k_y} \frac{{\bf d}_{2D}(\vec{k})}{|{\bf d}_{2D}(\vec{k})|}  \cdot {\bm \sigma} \Big],\nonumber
\end{align}
with $\ket{\psi_{{\bf k},-}}$ indicating the negative-energy
eigenstates. 
The averaging over vertical momentum, 
using the expression for the vector ${\bf d}_{2D}$,
can be written
\begin{align}
    \frac{1}{L} \sum_{ky} \frac{{\bf d}_{2D}(\vec{k})}{|{\bf d}_{2D}(\vec{k})|} 
    &= \big(\sin k_x,0,m-\cos k_x \big) {\cal F}(k_x)  \nonumber \\
    & -     
      \big(0,0,1\big) \frac{1}{L} \sum_{ky}  \frac{t_y \cos k_y}{|{\bf d}_{2D}(\vec{k})|}\nonumber, 
\end{align}
in which ${\cal F}(k_x)=(1/L)\sum_{k_y}1/|\vec d_{2D}(\vec{k})|$.
It is clear that the average of second component of the vector
identically vanishes due to its anti-symmetry under $k_y\to -k_y$.
The second line can also simply be absorbed inside the mass term $m$ as a renormalization, 
$$\delta m (k_x) = \frac{1}{{\cal F}(k_x)}\frac{t_y}{L} \sum_{k_y}\frac{\cos k_y}{|\vec d_{2D}(\vec{k})|} .$$
Putting altogether, the correlation matrix of the 1D subsystem
takes the following form
\begin{align}
      {\mathbfcal C}(k_x)  =
     \frac{1}{2}\big[ {\mathbbm 1} - 
    {\cal F}(k_x) \: \vec d_{1D}(k_x)
      \cdot {\bm \sigma} \big], \label{eq:QWZ-subsys-corr}
\end{align}
with
\begin{align}
    \vec d_{1D}(k_x)=\big[\sin k_x, 0, m-\delta m(k_x)-\cos k_x\big].\label{eq:1D}
\end{align}

\begin{figure}[t!]
\includegraphics[width=0.99\linewidth]{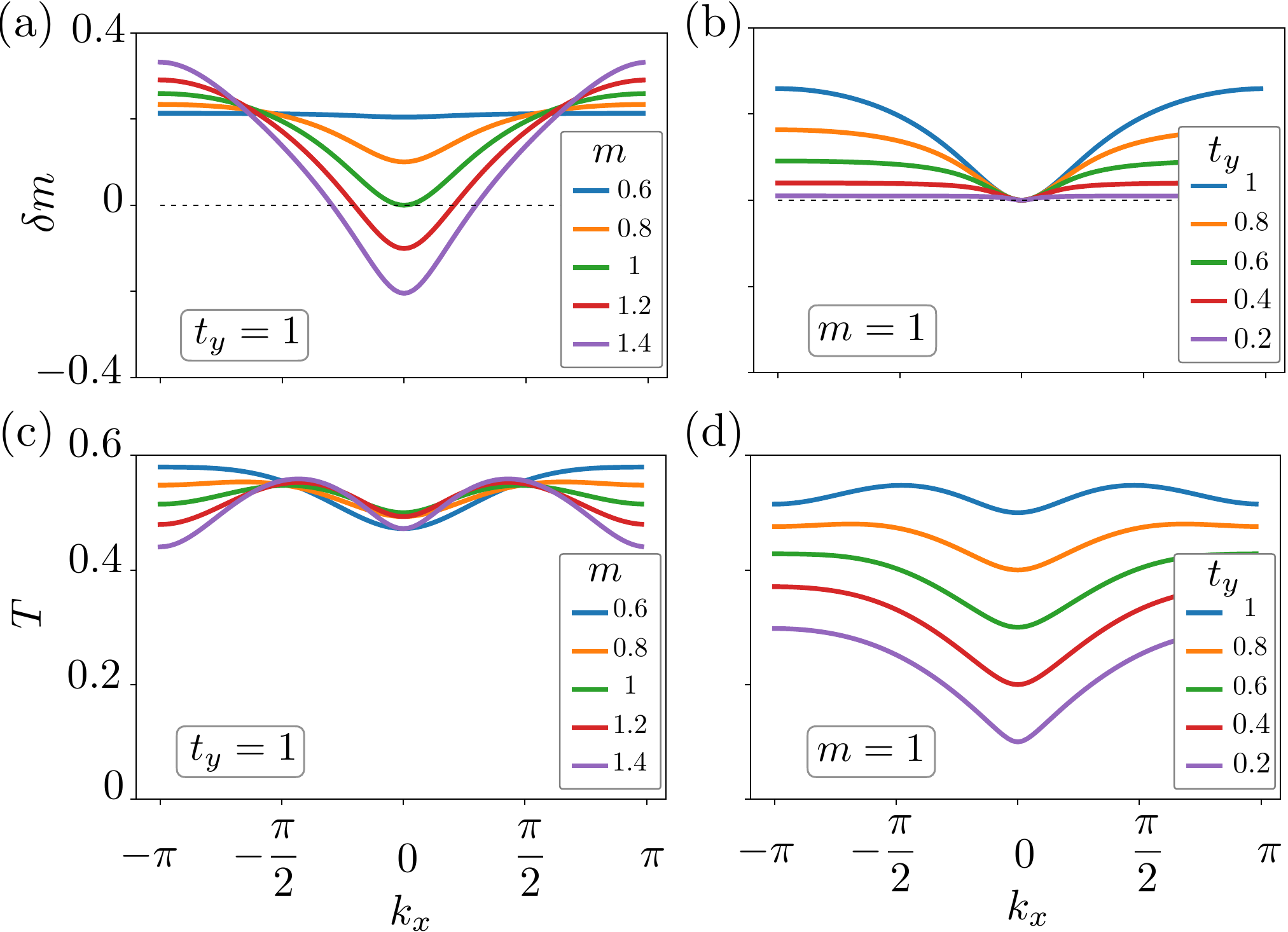} 
\caption{Entanglement temperature and mass renormalization in the QWZ model: (a) and (b) panels
show the dependence of mass renormalization $\delta m$ on momentum $k_x$ for different values of $m$ and $t_y$, respectively. As discussed in the text, $\delta m$ identically vanishes at the gap closing point when $m=1$ and $k_x=0$.
(c) and (d) panels show variation of effective temperature with momentum.
\label{fig:T_eff-vs-k_x}}
\end{figure}

The 1D ESH determining the thermal state is given by Eq.~\eqref{eq:1D} which has the form of $H_{2D}$ with vanishing transverse hopping $t_y=0$ and renormalized mass $m+\delta m(k_x)$.
In other words, the 1D subsystem embedded in the 2D QWZ model
has the same static properties as a vertically decoupled 1D chain 
with just a renormalized mass, and subjected to a temperature as will be elucidated more clearly in the following.
The dependence of the mass renormalization term $\delta m$ on momentum for different values of $m$ and $t_y$ is shown in Figs. ~\ref{fig:T_eff-vs-k_x}(a) and (b), respectively. 
Intriguingly, the mass renormalization vanishes identically when $m=1$ and $k_x=0$, which can be also deduced from the
mass renormalization expression by noticing that
at this particular point we have $\vec d_{2D}(k_x=0,k_y) =t_y(0,\sin k_y,-\cos k_y)$,
thereby $\delta m (k_x=0) |_{m=1}=0$.
But since $m=1$ corresponds to the gap-closing point of the 1D model 
with $\tilde{\vec d}_{1D} = \big( \sin k_x, 0 ,m-\cos k_x \big)$,
above observation implies implies that the gap-closing point of the ESH given by Eq. \eqref{eq:1D} at $m=1$ is not affected by $\delta m$.
As shown in Appendix \ref{appendix:gap-closing}, this behavior is not limited to the simple model with just nearest-neighbor hopping.
We also see that $\delta m$ is suppressed by decreasing the lateral hopping and vanishes when $t_y\to 0$ as expected.

The result \eqref{eq:QWZ-subsys-corr} can be recast into a manifestly thermal form  as
\begin{equation}\nonumber
    {\mathbfcal C}(k_x) = \frac{1}{2}\sum_{\eta=\pm1}\Big( {\mathbbm 1} + \eta 
   \frac{{\bf d}_{1D}}{|{\bf d}_{1D}|}\cdot {\bm \sigma}
   \Big)n_F(\eta|{\bf d}_{1D}|),
\end{equation}
where the temperature is obtained from Eq.~\eqref{eq:T_eff} as
\begin{equation}
    T(k_x)= \frac{|{\bf d}_{1D}|}{2\:{\rm arctanh}(|{\bf d}_{1D}|{\cal F})}. \nonumber
\end{equation}
This temperature is plotted in Fig.~\ref{fig:T_eff-vs-k_x}(c) for various values of mass $m$ and in \ref{fig:T_eff-vs-k_x}(d) for different values of transverse hopping $t_y$. As seen in Fig.~\ref{fig:T_eff-vs-k_x}(d), the scale of the temperature is set by transverse hopping $t_y$ as expected. The temperature has a weak dependence on momentum, especially around $m\sim1$ which corresponds to the gap closing of the effective 1D model \eqref{eq:1D}. 
Now, remembering that at $m=1$ the parent 2D system is gapped,
this has an interesting consequence that
the reduced density matrix of a 1D subsystem for $m=1$ matches that of a \emph{gapless} system at a very high temperature, even though we started from a gapped 2D system in its ground state. This result stems from the fact that, as we have noticed earlier, the ESH is equivalent to decoupled 1D chains accompanied with a mass renormalization which itself vanishes at the gap-closing point of the decoupled chains. Moreover, as we will see in the next section, a similar behavior is also revealed in higher dimensions, which makes this result quite profound,
especially noticing that the inclusion of further hoppings terms 
may result in the same final result (Appendix \ref{appendix:gap-closing}). 
In Sec. \ref{sec:dim_red}, we will elaborate more on the physical 
reasons behind the appearance of gapless subsystems of gapped systems, in a broader sense.

\begin{figure}[t!]
\includegraphics[width=0.65\linewidth]{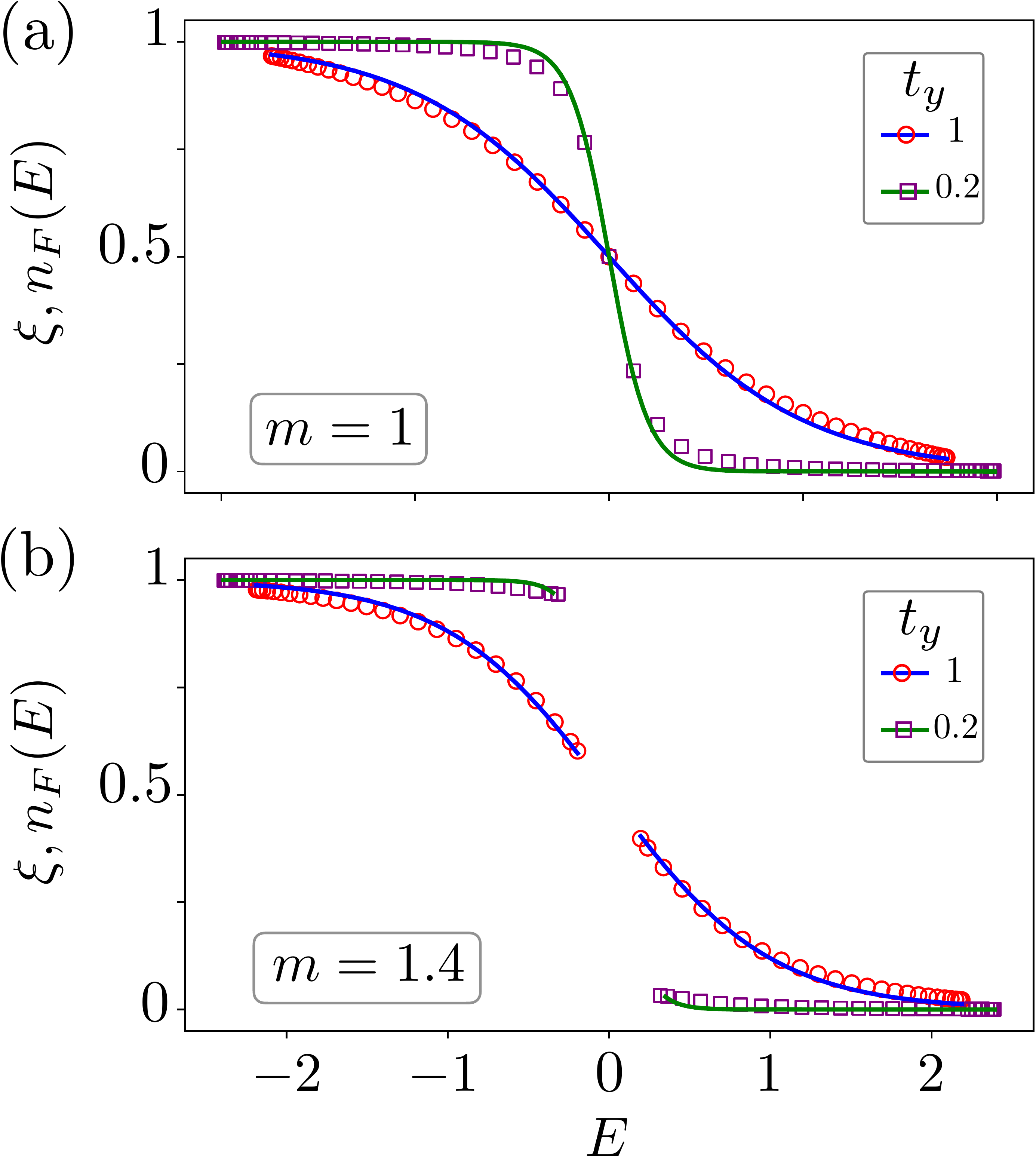} \leftskip1.1cm
\caption{Thermal population of a 1D subsystem in the QWZ model: (a) Correlation matrix spectra for the 1D subsystem (shown by circles/squares) and a thermal system (shown by lines) 
for $m=1$ and two different values of lateral hopping.
(b) Similar results for $m=1.4$.
In both panels, the thermal correlation spectrum is given by the Fermi-Dirac distribution $n_F(E)$  at temperatures
$T=0.5$ (for $t_y=1$), $T=0.1$ (for $t_y=0.2$), and plotted as a function of the ESH energy $E=\pm|{\bf d}_{1D}(k_x)|$ .
\label{fig:fermi}}
\end{figure}

Thermalization is further confirmed in Fig.~\ref{fig:fermi}, which shows comparison between the exact correlation matrix eigenvalues (denoted by $\xi$) and the corresponding thermal model with constant temperature. The correlation matrix eigenvalues provide the occupation probabilities of the subsystem states and are given by the Fermi-Dirac distribution at finite temperature. As seen in Fig.~\ref{fig:fermi}(a), the constant temperature Fermi-Dirac distribution essentially reproduces the exact results. Away from $|m|=1$, the ESH \eqref{eq:1D} is gapped, as indicated by the correlation matrix spectrum in Fig.~\ref{fig:fermi}(b). In Sec.~\ref{sec:experiment} we discuss how the entanglement-induced thermal state and the gapless subsystems can be observed through experimentally measurable fluctuations.

\subsection{Example II: Dirac models with linear dispersion}

We now show that the entanglement-temperature mapping of lower-dimensional subsystems becomes simple for continuum Dirac models with linear dispersion in arbitrary spatial dimensions. For simplicity, we consider the two-dimensional case but generalization to higher dimensions is straightforward. Let's consider Hamiltonian $H_{2D}= \vec d_{2D}(\vec{k})\cdot{\bm\sigma}$ with $ \vec d_{2D}(\vec{k})=(k_x, k_y, m)$ representing a 2D massive Dirac Hamiltonian. Adapting the formulas \eqref{eq:ds} and \eqref{eq:F} derived for lattice systems to continuum,  we obtain the effective 1D Hamiltonian of the corresponding thermal system as $ \vec d_{1D}(\vec{k})=(k_x, 0, m)$ and
\begin{equation}\nonumber
{\cal F}=
\frac{1}{2\Lambda}\int_{-\Lambda}^{\Lambda}\frac{dk_y}{\sqrt{k_x^2+k_y^2+m^2}}\approx\frac{1}{2\Lambda}\ln\big( \frac{4\Lambda^2}{k_x^2+m^2}\big).
\end{equation}
Here, a finite high-energy cutoff $\Lambda$ is required to avoid 
logarithmic divergence of the integral, and the final results is justified by assuming $\Lambda\gg \sqrt{{k_x^2+m^2}}$. According to  Eq.~\eqref{eq:T_eff}, 
the effective temperature reads 
\begin{equation}\label{eq:ref}
    T=\frac{\Lambda}{2\ln(2\Lambda/\sqrt{k_x^2+m^2})}\approx\frac{\Lambda}{2\ln(2\Lambda/m)}
\end{equation}
which means that it is only weakly-dependent on momentum and becomes constant in the small-momentum limit $k_x\ll m \ll \Lambda$. Similarly for higher dimensions, the effective $D_s$-dimensional Hamiltonian determining the reduced density matrix in \eqref{eq:reduced} is given by $H_{D}(\vec{k}_s)$ and the scale of the effective temperature is set by the cutoff scale $\Lambda$. The result Eq.~\eqref{eq:ref} shows that, for the temperature mapping to apply at small momentum, it is necessary to have a finite mass $|m|>0$ to avoid infrared divergences.

\subsection{ESH versus Bisognano-Wichmann Hamiltonian} \label{subsec-BHvsESH}
Above we noted the difference between the entanglement Hamiltonian and the ESH for lower-dimensional subsystems. Here we emphasize that the entanglement temperature and the ESH found in our work neither follow nor are consistent with the well-known Bisognano-Wichmann (BW) theorem which has enjoyed renewed interest recently. This theorem states that the entanglement Hamiltonian 
for a half-partitioning of Lorentz-invariant systems exactly follows the system Hamiltonian with just an additional spatially-varying prefactor besides the local Hamiltonian density of the original system \cite{BW1975,dalmonte2018quantum}. 
The first obvious difference is that unlike BW Hamiltonian, the ESH need not contain any position dependence since our subsystems are of lower dimension and are (lattice) translation invariant themselves. 
The second and even more important difference is that
the ESH does not follow the system Hamiltonian
as we have seen explicitly for prototype examples. In fact, the ESH can have completely different spectrum and physical behaviour than the original Hamiltonian. 
This is reflected in the surprising physical effect that the reduced density matrices of lower-dimensional subsystems of a gapped parent system can mimic that of a gapless system. Moreover, our thermal mapping is mathematically exact, and we find a renormalization of the Hamiltonian parameters  -- for instance, the mass parameter $m$ of Dirac models -- in the ESH, which is not consistent with the BW form of the entanglement Hamiltonian.

On top of the above distinctions, the temperature associated to the ESH and the renormalization of the parameters is controllable by adjusting the lateral hopping strength as seen above. This possibility to physically disentangle the notion of temperature and the ESH is absent
in the BW framework since relying just on the standard notion of entanglement Hamiltonian and the BW form, we always suffer from indistinguishable dual interpretations: (a) a uniform Hamiltonian accompanied by spatially-varying effective temperature,
(b) spatially varying BW Hamiltonian and a constant effective temperature.
Not to mention that we can have even further mathematically valid choices between the two above limiting interpretations.

Even if one attempted to settle the ambiguity of disentangling physically-relevant temperature by simply regarding the BW form of the reduced density matrix in terms of a constant Hamiltonian and spatially varying temperature profile, one would run into serious practical problems when trying to experimentally confirm the temperature profile.
To probe the reduced density matrix, one would need probe the reduced system as a whole. But since postulated effective entanglement temperature profile is strongly spatially dependent, one would need to probe the reduced system also locally to confirm the temperature profile. This leads to a dichotomy that one would simultaneously need to observe the whole system as well as probe its local properties. Thus, experimentally measuring any spatially-varying temperature associated with the BW theorem is deeply problematic in ways that highlights its different nature with ordinary temperature profiles.

\subsection{ Thermal subsystems from vacuum entanglement vs. genuine thermal states}
Since a density matrix encodes the full information of the state of a system at a given moment in time, all single-time expectation values and subsystem observables obtained from the density matrix \eqref{eq:reduced} will coincide with those of a $D_s$-dimensional system with Hamiltonian $H_{D_s}$ at finite temperature. However, it is clear that the entanglement-induced effective thermal subsystems exhibit crucial departures from true thermal states. In general, thermal systems emit thermal radiation and perturb their environment by thermal fluctuations. Since the full $D$-dimensional system \eqref{eq:dirac} is in the ground state, it is obviously impossible to extract net energy from any of its subsystems. Thus, contrary to naive expectations, the static thermal mapping \eqref{eq:reduced} does not imply that lower-dimensional subsystems would inherit all the properties of thermal states. 

To further quantify the above stated limitations, one can consider time-dependent generalization of the correlation operator  Eq.~\eqref{eq:subsys-general-corr}
\begin{equation}\label{eq:subsys-time-corr}
	\hat{\mathbfcal C}^{\rm sub}({\bf k}_{s};t)= \frac{1}{L^{D-D_s}}
	\sum_{{\rm filled}~\nu,{\bf k}_{\perp}}
	e^{-iE_{\nu{\bf k}}t}|\psi_{\nu{\bf k}}\rangle \langle \psi_{\nu{\bf k}}|,
\end{equation}
which depends on the full energy spectrum (excitations) $E_{\nu{\bf k}}$ of the higher dimensional parent 
system. Since the density matrix \eqref{eq:reduced} contains only the ground state information, it is insufficient in obtaining time and frequency-dependent correlations necessary to establish many standard properties of thermal systems such as the fluctuation-dissipation theorem. At very short times compared to the inverse of bandwidth ($t\ll1/\Delta E$), we only need to retain the ground state in Eq.~\eqref{eq:subsys-time-corr}
and the short time correlations from the static density matrix \eqref{eq:reduced}.
However, when $t\gtrsim 1/\Delta E$,
the full spectrum and excited states of the higher dimensional parent system become relevant to the subsystem properties, breaking the correspondence to genuinely thermal systems. As a consequence, 
the entanglement-induced thermal subsystems do not emit thermal radiation, display Johnson-Nyquist noise or obey fluctuations-dissipation relations. Furthermore, we cannot expect thermal signatures 
in any linear-response quantities 
as they also depend on frequency-resolved correlations (spectral functions of the full system). Thus, in sharp contrast to single-time expectation values, the properties sensitive to temporal correlations behave drastically differently from true thermal systems.

\section{Thermal subsystems and the table of topological insulators} \label{sec:dim_red}

The entanglement-temperature mapping for lower-dimensional subsystems has particularly interesting implications for topological materials. These materials can be arranged into a periodic table in terms of symmetry class and dimensionality, which repeats itself in every 8 dimensions \cite{Schnyder2008,kitaev2009}. The topological classes of adjacent dimensionality are connected through Bott periodicity, which maps a topological system in $d$ dimensions to one in $d+1$ dimensions with the same topological invariant by adding or removing chiral symmetry. Typically, this is used to establish connections between different physical systems, e.g. between one-dimensional chains and the scattering invariant of two-dimensional systems \cite{fulga2012}. Alternatively, one can introduce additional variables describing synthetic dimensions to carry out quantized pumping, which can also be realized experimentally \cite{Zilberberg2018,Lohse2018}. 

Since topological phases at different dimensions have Dirac Hamiltonian representatives, we can apply the entanglement-temperature mapping to study them. We show that the reduced density matrices of lower-dimensional subsystems have thermal form with respect to ESHs that exhibit the same topological classification as the table of topological insulators. By carrying out different subsystem measurements, the dimensional reduction actually becomes observable in a single physical system. Furthermore, we will illustrate the general pattern of how a hot gapless $D_s$-dimensional subsystem emerges from a $D$-dimensional gapped vacuum state, as pointed out in Subsec.~\ref{subsecChern}.

\subsection{Dimensional reduction from the 4D parent state}     
To demonstrate the connection between the thermal subsystem entanglement spectra and the dimensional hierarchy of topological materials, we explicitly derive lower-dimensional reduced density matrices of the 4D quantum Hall state
\cite{zhang2001,qi-zhang2008}. This model is widely known as the \emph{parent} Hamiltonian for descendants topological states using the standard dimensional reduction procedure \cite{qi-zhang2008,Schnyder2008,kitaev2009,qi-zhang-2011,Ryu2016}.
The lattice version of this model can be written in the form \eqref{eq:dirac} with a 5-component vector
\begin{align}\label{eq:4D-QH-model}
    {\bf d}_{4D}({\bf k})  = (m-\sum_{i=1}^4\cos k_i)\,\hat{\bf e}_0+\sum_{i=1}^4\sin k_i\,
    {\bf e}_i \:,
\end{align}
which depends on 4D momentum $\vec{k}$.  Here, we can introduce a basis  where the five $\Gamma$ matrices are given by
${\bm \Gamma} = (\tau_z\otimes\sigma_0,\tau_y\otimes\sigma_x,
     \tau_y\otimes\sigma_y,\tau_y\otimes\sigma_z,\tau_x\otimes\sigma_0)$.
The spectrum of the Hamiltonian possesses a pair of twofold degenerate bands with energies $\varepsilon_{\pm}({\bf k})=\pm|{\bf d}({\bf k})|$. 
Unlike the 2D Chern insulator which explicitly breaks time-reversal symmetry (TRS), the corresponding 4D model has a time-reversal symmetry ${\cal T} H_{\rm{4D}}({\bf k})  {\cal T}^{-1} = H_{\rm{4D}}(-{\bf k})$
with time-reversal operator ${\cal T}=i \tau_z \otimes \sigma_y \,{\cal K}$ based on the above choice for $\Gamma$ matrices. 
Hence, the Hamiltonian \eqref{eq:4D-QH-model} belongs to the symmetry class AII in the periodic table of the topological insulators. Nonetheless, since the topological classification of 4D topological phases in class AII and A coincide, we can equally consider the same model as a parent Hamiltonian in class A by adding a small TRS breaking term. Then, according to the Bott periodicity
depending on the symmetry class of the parent 4D system, we obtain two different generations of topological phases in lower dimensions belonging to different symmetry classes as summarized in Table \ref{table1}. 
\begin{table}
\caption{Dimensional reduction: symmetry classes and phase boundaries of 4D parent system and lower dimensional subsystems
	\label{table1} 
	}
\begin{ruledtabular}
\begin{tabular}{l @{} l @{} l}
Dimension & Symmetry class & gapless points\\
\hline
\vspace{-6pt}\\
	4D
	& 
	AII \:\:\qquad A
	& 
	$m_c=\pm4,\pm2,0$
	\\
	3D
	& 
	DIII \qquad AIII
	& 
	$m_c=\pm3,\pm1$
	\\
	2D
	& 
	D \,\quad\qquad A
	& 
	$m_c=\pm2,0$
	\\
	1D
	& 
	BDI \,\qquad AIII
	& 
	$m_c=\pm1$
	\\
    0D
	& 
	AI \:\:\:\:\qquad A
	& 
	$m_c=0$
	\\
\end{tabular}
\end{ruledtabular}
\end{table}
\par
Next, we consider lower dimensional subsystems of the 4D Hamiltonian  \eqref{eq:4D-QH-model}.
According to Eq.~\eqref{eq:sub-corr-Dirac}, for generalized Dirac Hamiltonians \eqref{eq:dirac}, the subsystem density matrix is determined by the effective Hamiltonian obtained by averaged ${\bf d}$-vector over the $4-D_s$ transverse momenta. Thus, the ESH is determined by Eqs.~\eqref{eq:ds},\eqref{eq:F} and given by 
\begin{align}
    &{\bf d}_{D_s} = \big[m-\delta m_{D_s}({\bf k}_{D_s}) -\sum_{i=1}^d\cos k_i\big]\hat{\bf e}_0
     + \sum_{i=1}^{d} \sin k_i \hat{\bf e}_i,\label{eq:subsystems-d-vector}\\
    & \delta m_{D_s}  = \frac{1}{{\cal F}_{D_s}}\:
    \int\frac{dk_4\cdots dk_{D_s+1}}{(2\pi)^{4-D_s}}
    \: \frac{\cos k_4+\cdots+\cos k_{D_s+1}}{|{\bf d}_{\rm{4D}}|}. \nonumber
\end{align}
The entanglement temperature then follows from Eq~\eqref{eq:T_eff}. The gapless points of the ESHs, signifying possible topological phase boundaries, are given by the condition ${\bf d}_{D_s}=0$. This can only take place at the high symmetry points $Q_i$ of the subsystem Brillouin zone, 
where $\sin Q_i =0$ for $i=0,\cdots,d$. Hence, at different $Q$-points, the gap closing condition becomes ${\bf d}_{D_s}({\bf Q})=\big[m-\delta m_{D_s}({\bf Q}) -\sum_{i=1}^d\cos Q_i\big]\hat{\bf e}_0\equiv0$. 
Since the shifts in the mass vanish at high symmetry points ($\delta m_{D_s}({\bf Q})=0$) implied by \eqref{eq:subsystems-d-vector}, the critical values are then given by $m_c=\sum_{i=1}^d\cos Q_i$.

\begin{figure}[t!]
\includegraphics[width=0.99\linewidth]{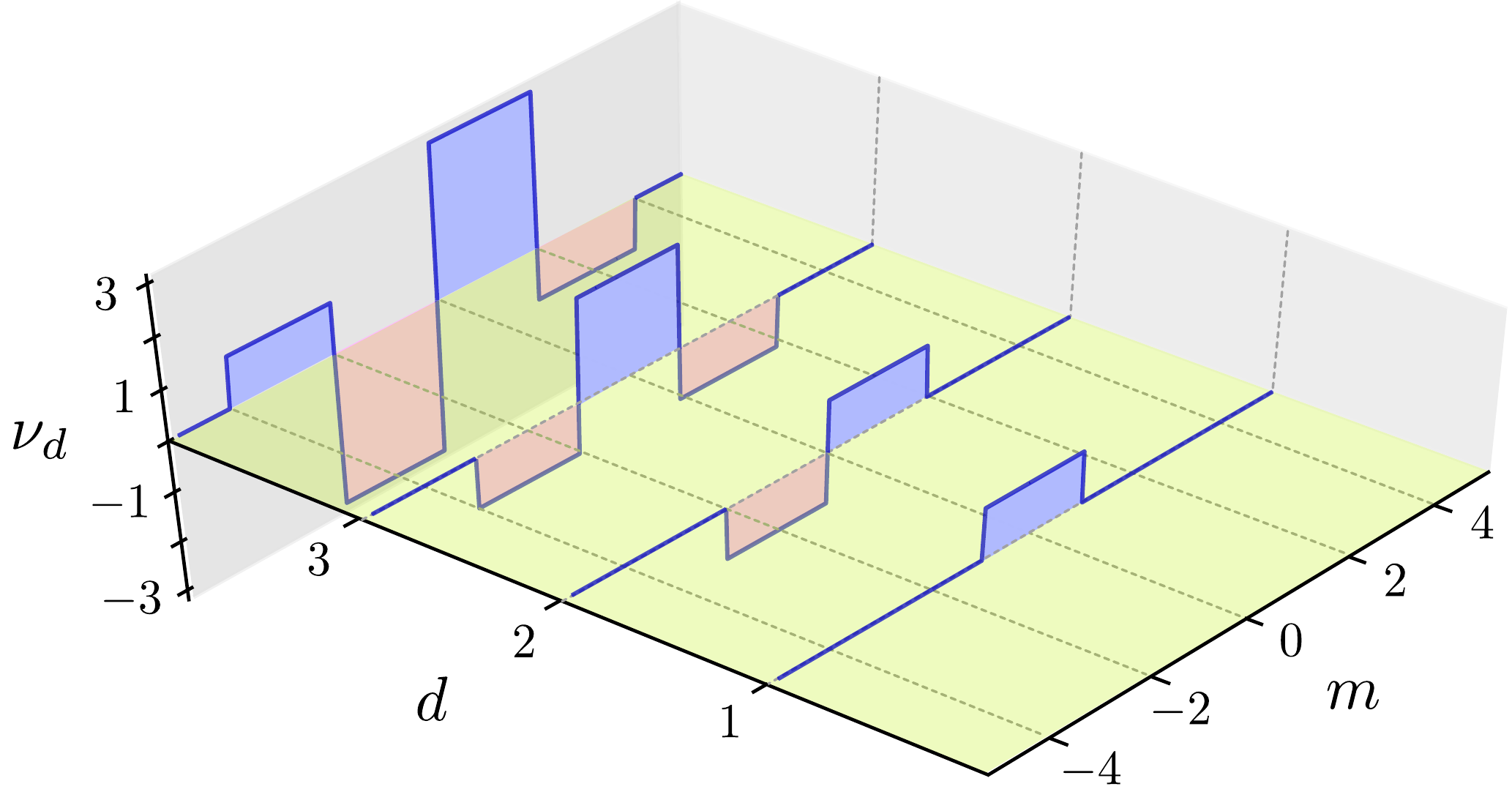} 
\caption{Topological invariant $\nu_d$ of the 4D QH model and its lower-dimensional effective subsystem Hamiltonians as a function of the band mass $m$. 
\label{fig:topo-inv-m}}
\end{figure}
\par

All of the descendent models as well as the parent systems have a ${\mathbb Z}$-classified topology
which is characterized by Chern and winding numbers in even and odd dimensions, respectively. This property holds irrespective of whether one regards the 4D parent state as belonging to class A or AII. For a Hamiltonian given in terms of $n+1$ different anticommuting Dirac matrices
and in $n$ spatial dimensions, the ${\mathbb Z}$ invariant has a generic form
\begin{equation}
    \nu_n = \frac{1}{S_n}\int d^n k \: \epsilon^{\mu_0\cdots\mu_n} \: \hat{d}_{\mu_0} 
    (\partial_{k_1}\hat{d}_{\mu_1})
    \cdots
    (\partial_{k_n}\hat{d}_{\mu_n}), \nonumber
\end{equation}
in terms of the mapping ${\hat{\bf d}}({\bf k})={\bf d}({\bf k})/|{\bf d}({\bf k})|$
from the $n$-dimensional Brillouin zone to the $n$-dimensional unit sphere \cite{footnote1}. The prefactor $S_n=2\pi^{(n+1)/2}/\Gamma[(n+1)/2]$ given in terms of gamma function, denotes the area of $n$-dimensional unit sphere.
The topological invariant for the 4D parent Hamiltonian and the lower dimensional entanglement Hamiltonians can be evaluated straightforwardly and the results are summarized in Fig. \ref{fig:topo-inv-m}. We observe that the topological invariant always changes at each gapless point and then
identically vanishes for $|m|>d$ (where $1\leq d\leq4$) indicating a trivial topological phase.
Particularly, we find that the ESHs have distinct topological landscapes with phase boundaries that move with the subsystem dimension $d$. Although this conclusion relies on the specific model \eqref{eq:4D-QH-model}, the gap closing pattern of the lower-dimensional ESHs is more general, as discussed in Appendix \ref{appendix:gap-closing}. At the critical point of the subsystem, the spectrum of the ESH actually describes a semimetal at finite temperature. This systematizes the observation in Subsec. \ref{subsecChern} that the lower-dimensional subsystems of a gapped system at zero temperature may actually appear as a metallic state at finite temperature. 
It also offers an intuitive explanation for the emergence of the gapless subsystems as follows. We have demonstrated that the process of tracing out the higher-dimensional complement to obtain the lower-dimensional ESH has analogous features with the usual process of dimensional reduction. 
In this process, the system parameters controlling the gap closings of lower-dimensional systems are generically renormalized. Likewise, the renormalization of parameters in the ESH, which is ultimately responsible for the emergence of hot gapless subsystems of zero-temperature insulators, can be regarded as reflection of what is expected from the conventional dimensional reduction. This argument applies to generic spatial dimension and also suggest it is largely insensitive to the details of the considered model. 
Therefore, from a physical point of view, the emergence of hot metallic subsystems from vacuum fluctuations of gapped systems is not limited to a specific model which makes it even more remarkable.

Finally, we note that the symmetries of the dimensional-reduced entanglement Hamiltonians are in agreement with the Bott periodicity of topological insulators. As can be seen in Eq.~\eqref{eq:subsystems-d-vector},
the $d$-dimensional entanglement Hamiltonian depends only on the first $d+1$ Dirac matrices.
As a consequence, 3D subsystem not only inherits the TRS from the parent 4D Hamiltonian, but also acquires a particle-hole symmetry (PHS) as ${\cal P} H_{\rm{3D}}({\bf k}) {\cal P}^{-1}=-H_{\rm{3D}}({\bf k}) $
with ${\cal P}=\tau_y\otimes \sigma_y\,{\cal K}$.
The presence of both TRS and PHS induce chiral symmetry ${\cal C}={\cal P}{\cal T}=\tau_x\otimes\sigma_0$, indicating that the 3D model belongs to the class DIII. With similar reasoning, one can figure out the symmetry classes of the lower-dimensional descendants as listed in Table. \ref{table1}. Thus, the entanglement-temperature mapping reflects the periodic table of topological insulators. In Appendix \ref{app:WTI}, we discuss how the reduced density matrix also reflects the topological properties of weak topological insulators.

\section{Experimental consequences} \label{sec:experiment}

The emergence of thermal states from the ground-state entanglement reflects the highly non-trivial nature of the quantum vacuum.  Although two special cases of this phenomenon, the Hawking and Unruh effects, have been known for half a century, the phenomenon has eluded experimental confirmation. The first experimental observation of thermal states from vacuum entanglement would be an outstanding achievement, bridging fundamental notions of quantum information, statistical physics, condensed matter physics, and high-energy physics. Here we propose a concrete setup to observe the vacuum thermalization within currently existing technology. The most natural setting for exploring our 
findings is ultracold atoms in optical lattices.
Such systems are considered ideal for quantum simulation for a wide variety of quantum phenomena due to their high level of control and accuracy \cite{Greiner2016quantum,Greiner2015measuring,Bloch2017quantum,brydges2019probing}.
Moreover, it has been previously established that these systems can realize various topological systems \cite{jotzu2014Haldane,aidelsburger2015measuring,goldman2016topological}. In particular, the Haldane model \cite{jotzu2014Haldane} can be represented as a massive two-band Dirac Hamiltonian \eqref{eq:dirac} and is directly relevant for our discussion. Moreover, the QWZ model studied in \ref{subsecChern} has already been realized in bosonic systems \cite{wu2016realization,Cooper2019RMP}. Thus, two-band Dirac systems are suitable candidates for experimental studies.

The entanglement-induced thermalization in these systems could be 
probed by comparing the entanglement-governed fluctuations 
in subsystems with genuinely thermal fluctuations.
This idea had been previously used to introduce an effective temperature for subsystems of a 1D spin system \cite{Eisler2006Fluctuations}.
But as we have thoroughly discussed in Subsec.~\ref{subsec-BHvsESH},
for such situations a unique unambiguous definition of entanglement temperature is almost impossible due to its position dependence. 
Nevertheless, using the fluctuations, it has been found that the effective temperature of subsystem vanishes as $T_{\rm eff} \propto \log L/L$ for large subsystem sizes \footnote{This result reflects another problem with defining entanglement temperature for subsystems with the same dimensionality as the system, because the resulting average temperature vanishes for very large sizes.}.
In sharp contrast, for lower dimensional subsystems,
we see that the entanglement temperature is enormous as it scales with the lateral hopping term and can be much higher than the real temperature in the experiment. As long as the real temperature is low compared to the hopping amplitudes, it has little effect on the outcome of the experiment. Moreover, 
the effective entanglement temperature can be easily controlled by varying the hopping amplitude out of the subsystem. Fluctuations of the subsystem observables match those of thermal systems at corresponding temperature, thus providing a feasible experimental signature to probe the entanglement-induced thermalization. 

\begin{figure}
    \centering
    \includegraphics[width=\linewidth]{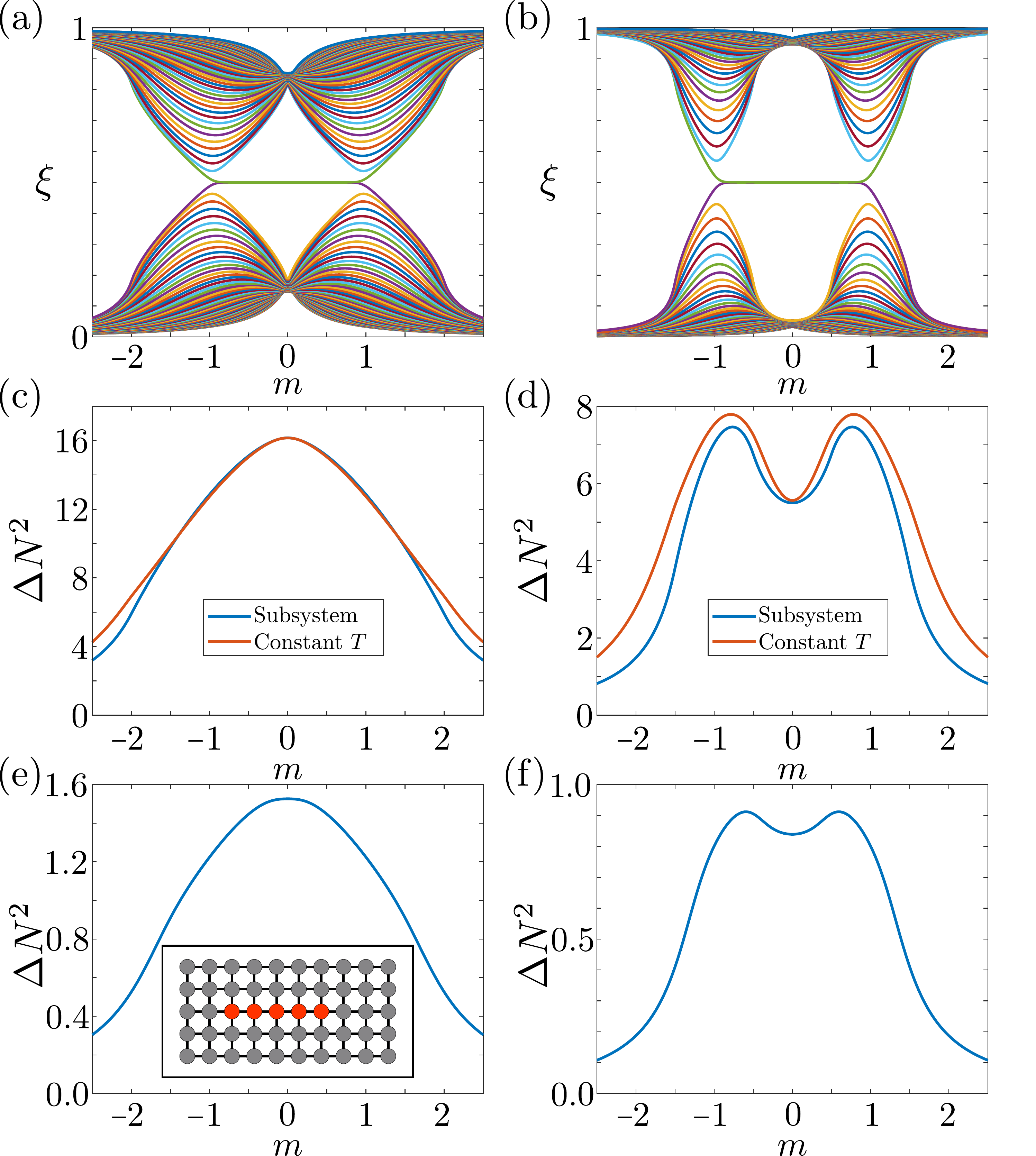}
    \caption{Correlation matrix spectra and particle number variances of finite 1D subsystems of a 2D QWZ model, with different orthogonal hoppings corresponding to different effective temperatures in the mapping to 1D systems. The red curve in the bottom figures also show the fluctuations obtained by use of Eq.~\eqref{eq:QWZ-subsys-corr}, but with a constant temperature simply equal to the mean over $k$ of the temperature given by Eq.~\eqref{eq:T_eff} (with the $k$-dependent  $T(k)$, the fluctuations would match exactly). (a) and (c): All hoppings of equal magnitude. (b) and (d): Orthogonal hoppings half the magnitude of hoppings parallel to the chain, corresponding to a lower effective temperature. 
    (e) Particle number variance of a subsystem of length $5$ in a total system of size $5\times10$ with open boundary conditions. Inset: schematic illustration of the system. (f) Same, but with $y$-directional hoppings half the magnitude of the ones in $x$ direction.}
    \label{fig:variances}
\end{figure}
\par

We illustrate the above recipe by studying the behavior of particle fluctuations in a 1D subsystem of the QWZ model studied in Subsec.~\ref{subsecChern}. The subsystem particle number operator is defined as $\hat{N}=\sum_{i}\hat{c}_i^{\dagger}\hat{c}_i$, where the summation is restricted to a chain in $x$ direction  embedded in the 2D lattice. The subsystem particle number fluctuations are quantified by their variance $\Delta N ^2=\langle \hat{N}^2 \rangle-\langle \hat{N} \rangle^2$.
We consider two different values for the lateral hopping $t_y/t_x=1$ and $t_y/t_x=0.5$, which translate to different effective temperatures. The corresponding correlation matrix spectra (or population probabilities) of the 1D subsystem are shown in Fig.~\ref{fig:variances}(a),(b). As seen in Figs.~\ref{fig:variances}(c) and (d), the comparison of the particle fluctuations as a function of the mass parameter in the subsystem with that of the associated 1D system at a constant temperature shows excellent agreement. When the effective temperature is reduced by decreasing the lateral hopping to $t_y/t_x=0.5$, the gap closing points of the ESH (and the correlation spectrum) at $m=\pm1$ become clearly visible through enhanced fluctuations signaled by the two peaks. The positions of these peaks, obscured by high effective temperature at $t_y/t_x=1$, do not coincide with the 2D gap closing points ($|m|=2$, $m=0$) but provide a smoking gun signature of our prediction that a 1D subsystem in the gapped 2D subsystem at zero temperature can appear as a hot gapless subsystem. 
Moreover, the fluctuations of conserved quantities can be exploited as an effective measure of entanglement entropy. 
The connection linking fluctuations and entanglement entropy,
which was previously considered in various studies especially in a transport context \cite{KlichLevitov2009,LeHur2010,LeHur2012},
has been put in a general framework by justifying the similarities
between entanglement entropy and variance of conserved subsystem observables
\cite{poyhonen2021observing}.
Thus, the agreement between the fluctuations, 
as shown in Fig. \ref{fig:variances}, implies that the entanglement entropy of the 1D subsystem corresponds to the thermodynamic entropy of a genuine 1D system at a constant temperature.

Importantly, the qualitative behavior of the fluctuations is preserved even at small system sizes accessible in current experiments. 
This is illustrated in Figs.~\ref{fig:variances}(e) and (f), where we show $\Delta N^2$ for a subsystem of length $5$ embedded in $5\times10$ array. Manipulating comparable lattice sizes are within reach of the current experimental techniques \cite{51_atom_simulator,53_atom_simulator,256_atom_simulator}. Moreover, a site-resolved measurement of particle number statistics, similar to what is needed in our proposal, has already been demonstrated in Ref.~\cite{Greiner2016quantum,Greiner2015measuring}. Thus, the thermal state arising from ground-state entanglement could be observed by realizing a two-band Dirac insulator and carrying out a site-resolved particle number measurement, both of which have been previously demonstrated in cold-atom experiments.

Finally, we emphasize the sharp distinction between our experimental proposal and a number of recent works with superficial similarities. First, the recent experiments simulating some aspects of the Hawking and Unruh effect \cite{steinhauer2016observation,munoz2019observation,hu2019Unruh}, unlike in our proposal, apply time-dependent driving to stimulate a thermal-like radiation in systems that are not described by a static thermal density matrix. In this sense, they do not constitute a demonstration of thermal states emerging from vacuum entanglement. Similarly, the purpose and the outcome of previous works simulating entanglement Hamiltonians \cite{dalmonte2018quantum,dalmoste2022,Kokail2022} are equally distinct from our theoretical proposal. The main purpose of these works is to artificially realize entanglement Hamiltonians for certain lattice models that follow, at least approximately, the BW Ansatz. As has been thoroughly discussed in Subsec.~\ref{subsec-BHvsESH},
the artificial simulation of entanglement Hamiltonians, while interesting in its own right, cannot be regarded as a confirmation of vacuum entanglement-induced thermalization, nor the cited works claim so. In our experimental scheme, we suggest to directly observe the subsystem particle number fluctuations that follow a thermal equilibrium form, thus directly revealing the thermal nature of the reduced system. In this definite sense, our proposal would indeed enable the first observation of vacuum-entanglement-induced effective temperature.

\section{Conclusion } 

In this work we identified a large class of quantum many-body systems, constituting of gapped Dirac fermions, in which entanglement of vacuum fluctuations give rise to a thermal density matrix in their lower-dimensional subsystems. We also showed that, remarkably, subsystems of a zero-temperature insulator may even appear as hot gapless systems. We proposed that the emergence of a thermal state from vacuum could be realistically observed, for the first time, in cold atom quantum simulators through thermal fluctuations. Direct experimental verification of an emergent thermal state from vacuum quantum fluctuations would be an outstanding achievement with ramifications in statistical physics, condensed-matter physics, high-energy physics, and quantum information.

\acknowledgements
The authors acknowledge the Academy of Finland project 331094 for support.

\appendix

\section{Gap closing in dimensionally reduced systems}
\label{appendix:gap-closing}

Let us assume we have a Dirac-type Hamiltonian
\begin{equation}
	H_k = \sum_{i=1}^{N} d_i(\vec k) \Gamma_i
\end{equation}
where $\Gamma_i$ are $2^n\times 2^n$ general gamma matrices that obey 
the anticommutation rule
$\lbrace \Gamma_i, \Gamma_j\rbrace = 2\delta_{ij}$.

Let us assume the system is $D$-dimensional and translationally invariant, and the subsystem considered for the $\mathcal C$ matrix is $(D-1)$-dimensional and likewise translationally invariant. As mentioned in the main text, the $\mathcal C$ matrix at $T = 0$ can then be written

\begin{equation}
	\mathcal C(k_\parallel)= \frac{1}{L}
	\sum_{{\rm filled}~\nu, k_\perp} |\psi_{\nu,{\bf k}}\rangle \langle \psi_{\nu,{\bf k}}|.
\end{equation}

In the above we have written $\vec k = (k_\parallel, k_\perp)$, where $k_\parallel$ are the momenta inside the subsystem and $k_\perp$ represents the single momentum component orthogonal to the subsystem. Due to the structure of the Hamiltonian, the above is equivalent to

\begin{equation}
	\mathcal C (\vec k_\parallel)= 
	{\mathbbm 1} 
	- \frac{1}{L}
	\sum_{k_\perp,i} \frac{d_i(\vec k)}{d(\vec k)}\Gamma_i \equiv {\mathbbm 1}
	- B(\vec k_\parallel)
\end{equation}

A gap closing corresponds to $B$ having zero-energy eigenvalues for some $\vec k_\parallel$. Let us now separate the part of the Hamiltonian that depends on the orthogonal momentum, i.e.
\begin{equation}
	d_i(\vec k) = h_i(k_\parallel) + f_i(k_\parallel,\vec k_\perp)
\end{equation}
In this way we have separated a lower-dimensional Hamiltonian expressed in terms of $h_i$:
\begin{equation}
	H_{D-1}(\vec k_\parallel) = \sum_{i=1}^{N} h_i(\vec  k_\parallel) \Gamma_i
\end{equation}
The functions $f$ may or may not be $\vec k_\parallel$-dependent (as e.g. would occur with diagonal hoppings).

Let us now consider a momentum $\vec k_\parallel = \vec q_0$ where $H_{D-1}$ has a gap closing, so that $\forall i:\ h_i(\vec q_0) = 0$. At this particular point, the $\Gamma$ components of $B$ take the form
\begin{equation}
	B(\vec q_0)_i = \frac{1}{L} \sum_{k_\perp} \frac{f_i(k_\perp,\vec q_0)}{\sqrt{\sum_j f_j(k_\perp,\vec q_0)^2}}.
\end{equation}
Let us first assume that for every $i$, the function $f_i(k_N,\vec q_0)$ is either even or odd around $k_\perp = 0$ (note that this does not need to be the case for a generic $\vec k_\parallel$). In this case, where it is odd $B_i$ vanishes, and where it is even it reduces to
\begin{equation}
	B(\vec q_0)_i = 2\frac{1}{L_N} \sum_{k_D < \pi} \frac{f_i(k_\perp,\vec q_0)}{\sqrt{\sum_j f_j(k_\perp,\vec q_0)^2}}.
\end{equation}
If now within this $k_\perp$ interval the numerator is odd and the denominator even around $\frac{\pi}{2}$, the whole sum vanishes. This will occur if for every $i$, and $0 \leq \epsilon \leq \frac{\pi}{2}$, $|f_i(\frac{\pi}{2} + \epsilon, \vec q_0)| = |f_i(\frac{\pi}{2} - \epsilon, \vec q_0)|$, and further $f_i(\frac{\pi}{2} + \epsilon,\vec q_0) = -f_i(\frac{\pi}{2} - \epsilon,\vec q_0)$ for those $i$ for which $f_i(k_\perp,\vec q_0)$ is even around $k_\perp = 0$. This will be the case e.g. for nearest-neighbour hopping on a lattice. If so, gap closings of $H_{D-1}$ immediately imply gap closings of $H_k$.

\section{Thermal lower-dimensional subsystems in weak topological insulators}
\label{app:WTI}
\begin{figure}[t!]
\includegraphics[width=0.99\linewidth]{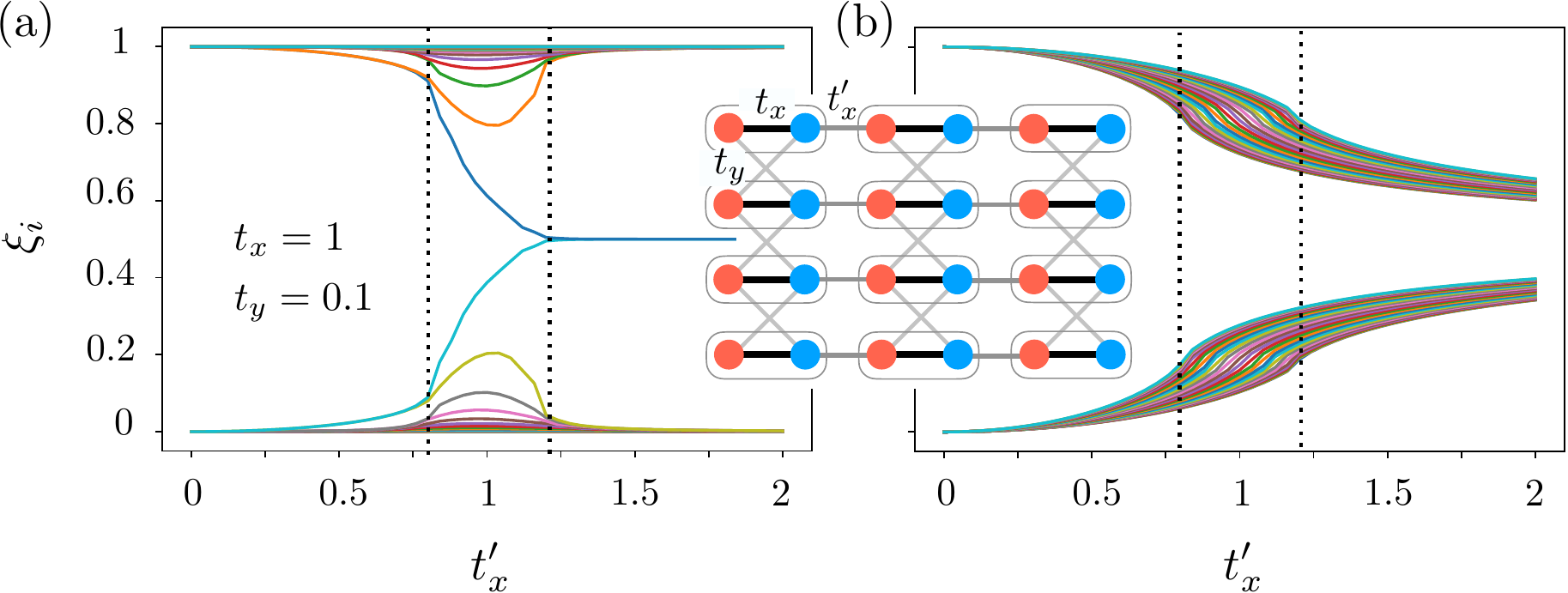}
\caption{Correlation matrix spectra for 1D subsystems of the 2D WTI.
The system size (number of unit cells) is $60\times60$ and the 1D sublattices have the length $L=30$. (a) and (b) indicate the spectra for a 1D open subsystem along $x$ and $y$, respectively. 
The dotted vertical lines are eye-guide for separating different phases i.e. trivial, gapless, and WTI, respectively.
\label{fig:WTI-ES}}
\end{figure}

Here we show how the lower-dimensional thermal subsystems reflect the topological properties of weak topological insulators. One way to construct a model for a 2D weak topological insulator (WTI) is
to consider a vertical stack of SSH chains with nearest-neighbor
unit cells coupled in the vertical direction as shown in the inset of Fig. \ref{fig:WTI-ES}.
The two-band Hamiltonian of the model can be written in the Dirac form 
$H_{2D}= {\bf d}_{\rm WTI}({\bf k})\cdot{\bm\sigma}$ with
\begin{equation}\label{eq:WTI}
    {\bf d}_{\rm WTI}({\bf k}) = \big(t_x + t'_x \cos k_x +2t_y \cos k_y,t'_x \sin k_x ,0\big).
\end{equation}
This Hamiltonian clearly satisfies the chiral symmetry as we have $\sigma_z {H} \sigma_z = - { H}$.
Hence, the topological characterization of the model is encoded in the weak indices
$\nu_j = -i\int  d^2{\bf k}/(2\pi)^2\: Q_{\bf k}^{-1} \,\partial^{\phantom\dag}_{k_j}Q_{\bf k}$,
which are based on the vertical averaging over the 1D winding number densities. Here, $Q_{\bf k}=d_x+id_y$ where $d_x$, $d_y$ denote the components of \eqref{eq:WTI}. 
Assuming $t_y>0$, phase diagram of this model consists of
a WTI with $(\nu_x,\nu_y)=(1,0)$ for $t'_x-t_x>2t_y$, a trivial phase
for $t_x-t'_x>2t_y$, and gapless (metallic) phase for $|t'_x-t_x|<2t_y$.

In the WTI phase, depending on the orientation of the reduced 1D subsystem with respect to $x$ and $y$ directions, the entanglement Hamiltonian is in the topological and the trivial phase, respectively. These distinction, which reflects the  the weak topological index of the parent 2D system, is illustrated in Fig. \ref{fig:WTI-ES}. The topology of the 1D ESH is easily obtained from the ${\bf d}$ vectors of the subsystems along the $x$ and $y$ directions,
\begin{align}
   d_{1x}(k)
   &= \big[t_x+t'_x\cos k + \delta_x t(k),t'_x\sin k,0\big]\\
   d_{1y}(k)&=  \big[t_x+2t_y\cos k + \delta_y t(k),0,0\big]
\end{align}
which are obtained from Eqs.~\eqref{eq:ds},\eqref{eq:F} by setting $\vec{k}_\perp=k_y, k_x$. This elucidates why the ESH for the subsystem along $y$ cannot have a topological phase since its winding number vanishes identically. In contrast, $d_{1x}(k)$ for the subsystem along the $x$ axis supports a finite winding number in the same regime as parent Hamiltonian \eqref{eq:WTI}.

\bibliography{biblio.bib}

\begin{thebibliography}{64}%
\makeatletter
\providecommand \@ifxundefined [1]{%
 \@ifx{#1\undefined}
}%
\providecommand \@ifnum [1]{%
 \ifnum #1\expandafter \@firstoftwo
 \else \expandafter \@secondoftwo
 \fi
}%
\providecommand \@ifx [1]{%
 \ifx #1\expandafter \@firstoftwo
 \else \expandafter \@secondoftwo
 \fi
}%
\providecommand \natexlab [1]{#1}%
\providecommand \enquote  [1]{``#1''}%
\providecommand \bibnamefont  [1]{#1}%
\providecommand \bibfnamefont [1]{#1}%
\providecommand \citenamefont [1]{#1}%
\providecommand \href@noop [0]{\@secondoftwo}%
\providecommand \href [0]{\begingroup \@sanitize@url \@href}%
\providecommand \@href[1]{\@@startlink{#1}\@@href}%
\providecommand \@@href[1]{\endgroup#1\@@endlink}%
\providecommand \@sanitize@url [0]{\catcode `\\12\catcode `\$12\catcode
  `\&12\catcode `\#12\catcode `\^12\catcode `\_12\catcode `\%12\relax}%
\providecommand \@@startlink[1]{}%
\providecommand \@@endlink[0]{}%
\providecommand \url  [0]{\begingroup\@sanitize@url \@url }%
\providecommand \@url [1]{\endgroup\@href {#1}{\urlprefix }}%
\providecommand \urlprefix  [0]{URL }%
\providecommand \Eprint [0]{\href }%
\providecommand \doibase [0]{http://dx.doi.org/}%
\providecommand \selectlanguage [0]{\@gobble}%
\providecommand \bibinfo  [0]{\@secondoftwo}%
\providecommand \bibfield  [0]{\@secondoftwo}%
\providecommand \translation [1]{[#1]}%
\providecommand \BibitemOpen [0]{}%
\providecommand \bibitemStop [0]{}%
\providecommand \bibitemNoStop [0]{.\EOS\space}%
\providecommand \EOS [0]{\spacefactor3000\relax}%
\providecommand \BibitemShut  [1]{\csname bibitem#1\endcsname}%
\let\auto@bib@innerbib\@empty
\bibitem [{\citenamefont {Kardar}(2007)}]{kardar2007book}%
  \BibitemOpen
  \bibfield  {author} {\bibinfo {author} {\bibfnamefont {Mehran}\ \bibnamefont
  {Kardar}},\ }\href@noop {} {\emph {\bibinfo {title} {Statistical physics of
  particles}}}\ (\bibinfo  {publisher} {Cambridge University Press},\ \bibinfo
  {year} {2007})\BibitemShut {NoStop}%
\bibitem [{\citenamefont {Deutsch}(1991)}]{Deutsch1991}%
  \BibitemOpen
  \bibfield  {author} {\bibinfo {author} {\bibfnamefont {J.~M.}\ \bibnamefont
  {Deutsch}},\ }\bibfield  {title} {\enquote {\bibinfo {title} {Quantum
  statistical mechanics in a closed system},}\ }\href {\doibase
  10.1103/PhysRevA.43.2046} {\bibfield  {journal} {\bibinfo  {journal} {Phys.
  Rev. A}\ }\textbf {\bibinfo {volume} {43}},\ \bibinfo {pages} {2046}
  (\bibinfo {year} {1991})}\BibitemShut {NoStop}%
\bibitem [{\citenamefont {Srednicki}(1994)}]{Srednicki1994}%
  \BibitemOpen
  \bibfield  {author} {\bibinfo {author} {\bibfnamefont {Mark}\ \bibnamefont
  {Srednicki}},\ }\bibfield  {title} {\enquote {\bibinfo {title} {Chaos and
  quantum thermalization},}\ }\href {\doibase 10.1103/PhysRevE.50.888}
  {\bibfield  {journal} {\bibinfo  {journal} {Phys. Rev. E}\ }\textbf {\bibinfo
  {volume} {50}},\ \bibinfo {pages} {888} (\bibinfo {year} {1994})}\BibitemShut
  {NoStop}%
\bibitem [{\citenamefont {Rigol}\ and\ \citenamefont
  {Srednicki}(2012)}]{Srednicki2012}%
  \BibitemOpen
  \bibfield  {author} {\bibinfo {author} {\bibfnamefont {Marcos}\ \bibnamefont
  {Rigol}}\ and\ \bibinfo {author} {\bibfnamefont {Mark}\ \bibnamefont
  {Srednicki}},\ }\bibfield  {title} {\enquote {\bibinfo {title} {Alternatives
  to eigenstate thermalization},}\ }\href {\doibase
  10.1103/PhysRevLett.108.110601} {\bibfield  {journal} {\bibinfo  {journal}
  {Phys. Rev. Lett.}\ }\textbf {\bibinfo {volume} {108}},\ \bibinfo {pages}
  {110601} (\bibinfo {year} {2012})}\BibitemShut {NoStop}%
\bibitem [{\citenamefont {Rigol}\ \emph {et~al.}(2008)\citenamefont {Rigol},
  \citenamefont {Dunjko},\ and\ \citenamefont
  {Olshanii}}]{rigol2008thermalization}%
  \BibitemOpen
  \bibfield  {author} {\bibinfo {author} {\bibfnamefont {Marcos}\ \bibnamefont
  {Rigol}}, \bibinfo {author} {\bibfnamefont {Vanja}\ \bibnamefont {Dunjko}}, \
  and\ \bibinfo {author} {\bibfnamefont {Maxim}\ \bibnamefont {Olshanii}},\
  }\bibfield  {title} {\enquote {\bibinfo {title} {Thermalization and its
  mechanism for generic isolated quantum systems},}\ }\href {\doibase
  10.1038/nature06838} {\bibfield  {journal} {\bibinfo  {journal} {Nature}\
  }\textbf {\bibinfo {volume} {452}},\ \bibinfo {pages} {854} (\bibinfo {year}
  {2008})}\BibitemShut {NoStop}%
\bibitem [{\citenamefont {Garrison}\ and\ \citenamefont
  {Grover}(2018)}]{Garrison2018}%
  \BibitemOpen
  \bibfield  {author} {\bibinfo {author} {\bibfnamefont {James~R.}\
  \bibnamefont {Garrison}}\ and\ \bibinfo {author} {\bibfnamefont {Tarun}\
  \bibnamefont {Grover}},\ }\bibfield  {title} {\enquote {\bibinfo {title}
  {Does a single eigenstate encode the full hamiltonian?}}\ }\href {\doibase
  10.1103/PhysRevX.8.021026} {\bibfield  {journal} {\bibinfo  {journal} {Phys.
  Rev. X}\ }\textbf {\bibinfo {volume} {8}},\ \bibinfo {pages} {021026}
  (\bibinfo {year} {2018})}\BibitemShut {NoStop}%
\bibitem [{\citenamefont {D'Alessio}\ \emph {et~al.}(2016)\citenamefont
  {D'Alessio}, \citenamefont {Kafri}, \citenamefont {Polkovnikov},\ and\
  \citenamefont {Rigol}}]{Polkovnikov2016ETH}%
  \BibitemOpen
  \bibfield  {author} {\bibinfo {author} {\bibfnamefont {Luca}\ \bibnamefont
  {D'Alessio}}, \bibinfo {author} {\bibfnamefont {Yariv}\ \bibnamefont
  {Kafri}}, \bibinfo {author} {\bibfnamefont {Anatoli}\ \bibnamefont
  {Polkovnikov}}, \ and\ \bibinfo {author} {\bibfnamefont {Marcos}\
  \bibnamefont {Rigol}},\ }\bibfield  {title} {\enquote {\bibinfo {title} {From
  quantum chaos and eigenstate thermalization to statistical mechanics and
  thermodynamics},}\ }\href {\doibase 10.1080/00018732.2016.1198134} {\bibfield
   {journal} {\bibinfo  {journal} {Adv. Phys.}\ }\textbf {\bibinfo {volume}
  {65}},\ \bibinfo {pages} {239} (\bibinfo {year} {2016})}\BibitemShut
  {NoStop}%
\bibitem [{\citenamefont {Deutsch}(2018)}]{Deutsch_2018}%
  \BibitemOpen
  \bibfield  {author} {\bibinfo {author} {\bibfnamefont {Joshua~M}\
  \bibnamefont {Deutsch}},\ }\bibfield  {title} {\enquote {\bibinfo {title}
  {Eigenstate thermalization hypothesis},}\ }\href {\doibase
  10.1088/1361-6633/aac9f1} {\bibfield  {journal} {\bibinfo  {journal} {Rep.
  Prog. Phys.}\ }\textbf {\bibinfo {volume} {81}},\ \bibinfo {pages} {082001}
  (\bibinfo {year} {2018})}\BibitemShut {NoStop}%
\bibitem [{\citenamefont {Biroli}\ \emph {et~al.}(2010)\citenamefont {Biroli},
  \citenamefont {Kollath},\ and\ \citenamefont {L\"auchli}}]{Lauchli2010}%
  \BibitemOpen
  \bibfield  {author} {\bibinfo {author} {\bibfnamefont {Giulio}\ \bibnamefont
  {Biroli}}, \bibinfo {author} {\bibfnamefont {Corinna}\ \bibnamefont
  {Kollath}}, \ and\ \bibinfo {author} {\bibfnamefont {Andreas~M.}\
  \bibnamefont {L\"auchli}},\ }\bibfield  {title} {\enquote {\bibinfo {title}
  {Effect of rare fluctuations on the thermalization of isolated quantum
  systems},}\ }\href {\doibase 10.1103/PhysRevLett.105.250401} {\bibfield
  {journal} {\bibinfo  {journal} {Phys. Rev. Lett.}\ }\textbf {\bibinfo
  {volume} {105}},\ \bibinfo {pages} {250401} (\bibinfo {year}
  {2010})}\BibitemShut {NoStop}%
\bibitem [{\citenamefont {Bombelli}\ \emph {et~al.}(1986)\citenamefont
  {Bombelli}, \citenamefont {Koul}, \citenamefont {Lee},\ and\ \citenamefont
  {Sorkin}}]{Sorkin1986}%
  \BibitemOpen
  \bibfield  {author} {\bibinfo {author} {\bibfnamefont {Luca}\ \bibnamefont
  {Bombelli}}, \bibinfo {author} {\bibfnamefont {Rabinder~K.}\ \bibnamefont
  {Koul}}, \bibinfo {author} {\bibfnamefont {Joohan}\ \bibnamefont {Lee}}, \
  and\ \bibinfo {author} {\bibfnamefont {Rafael~D.}\ \bibnamefont {Sorkin}},\
  }\bibfield  {title} {\enquote {\bibinfo {title} {Quantum source of entropy
  for black holes},}\ }\href {\doibase 10.1103/PhysRevD.34.373} {\bibfield
  {journal} {\bibinfo  {journal} {Phys. Rev. D}\ }\textbf {\bibinfo {volume}
  {34}},\ \bibinfo {pages} {373--383} (\bibinfo {year} {1986})}\BibitemShut
  {NoStop}%
\bibitem [{\citenamefont {Srednicki}(1993)}]{Srednicki1993}%
  \BibitemOpen
  \bibfield  {author} {\bibinfo {author} {\bibfnamefont {Mark}\ \bibnamefont
  {Srednicki}},\ }\bibfield  {title} {\enquote {\bibinfo {title} {Entropy and
  area},}\ }\href {\doibase 10.1103/PhysRevLett.71.666} {\bibfield  {journal}
  {\bibinfo  {journal} {Phys. Rev. Lett.}\ }\textbf {\bibinfo {volume} {71}},\
  \bibinfo {pages} {666} (\bibinfo {year} {1993})}\BibitemShut {NoStop}%
\bibitem [{\citenamefont {Frolov}\ and\ \citenamefont
  {Fursaev}(1998)}]{Frolov1998}%
  \BibitemOpen
  \bibfield  {author} {\bibinfo {author} {\bibfnamefont {V~P}\ \bibnamefont
  {Frolov}}\ and\ \bibinfo {author} {\bibfnamefont {D~V}\ \bibnamefont
  {Fursaev}},\ }\bibfield  {title} {\enquote {\bibinfo {title} {Thermal fields,
  entropy and black holes},}\ }\href {\doibase 10.1088/0264-9381/15/8/001}
  {\bibfield  {journal} {\bibinfo  {journal} {Class. Quantum Gravity}\ }\textbf
  {\bibinfo {volume} {15}},\ \bibinfo {pages} {2041--2074} (\bibinfo {year}
  {1998})}\BibitemShut {NoStop}%
\bibitem [{\citenamefont {Solodukhin}(2011)}]{solodukhin2011review}%
  \BibitemOpen
  \bibfield  {author} {\bibinfo {author} {\bibfnamefont {Sergey~N}\
  \bibnamefont {Solodukhin}},\ }\bibfield  {title} {\enquote {\bibinfo {title}
  {Entanglement entropy of black holes},}\ }\href {\doibase
  10.12942/lrr-2011-8} {\bibfield  {journal} {\bibinfo  {journal} {Living
  Reviews in Relativity}\ }\textbf {\bibinfo {volume} {14}},\ \bibinfo {pages}
  {8} (\bibinfo {year} {2011})}\BibitemShut {NoStop}%
\bibitem [{\citenamefont {Ryu}\ and\ \citenamefont
  {Takayanagi}(2006)}]{ryu2006}%
  \BibitemOpen
  \bibfield  {author} {\bibinfo {author} {\bibfnamefont {Shinsei}\ \bibnamefont
  {Ryu}}\ and\ \bibinfo {author} {\bibfnamefont {Tadashi}\ \bibnamefont
  {Takayanagi}},\ }\bibfield  {title} {\enquote {\bibinfo {title} {Aspects of
  holographic entanglement entropy},}\ }\href {\doibase
  10.1088/1126-6708/2006/08/045} {\bibfield  {journal} {\bibinfo  {journal} {J.
  High Energy Phys.}\ }\textbf {\bibinfo {volume} {2006}},\ \bibinfo {pages}
  {045} (\bibinfo {year} {2006})}\BibitemShut {NoStop}%
\bibitem [{\citenamefont {Casini}\ and\ \citenamefont
  {Huerta}(2009)}]{Casini2009}%
  \BibitemOpen
  \bibfield  {author} {\bibinfo {author} {\bibfnamefont {H}~\bibnamefont
  {Casini}}\ and\ \bibinfo {author} {\bibfnamefont {M}~\bibnamefont {Huerta}},\
  }\bibfield  {title} {\enquote {\bibinfo {title} {Entanglement entropy in free
  quantum field theory},}\ }\href {\doibase 10.1088/1751-8113/42/50/504007}
  {\bibfield  {journal} {\bibinfo  {journal} {J. Phys. A}\ }\textbf {\bibinfo
  {volume} {42}},\ \bibinfo {pages} {504007} (\bibinfo {year}
  {2009})}\BibitemShut {NoStop}%
\bibitem [{\citenamefont {Osterloh}\ \emph {et~al.}(2002)\citenamefont
  {Osterloh}, \citenamefont {Amico}, \citenamefont {Falci},\ and\ \citenamefont
  {Fazio}}]{osterloh2002scaling}%
  \BibitemOpen
  \bibfield  {author} {\bibinfo {author} {\bibfnamefont {Andreas}\ \bibnamefont
  {Osterloh}}, \bibinfo {author} {\bibfnamefont {Luigi}\ \bibnamefont {Amico}},
  \bibinfo {author} {\bibfnamefont {Giuseppe}\ \bibnamefont {Falci}}, \ and\
  \bibinfo {author} {\bibfnamefont {Rosario}\ \bibnamefont {Fazio}},\
  }\bibfield  {title} {\enquote {\bibinfo {title} {Scaling of entanglement
  close to a quantum phase transition},}\ }\href {\doibase 10.1038/416608a}
  {\bibfield  {journal} {\bibinfo  {journal} {Nature}\ }\textbf {\bibinfo
  {volume} {416}},\ \bibinfo {pages} {608} (\bibinfo {year}
  {2002})}\BibitemShut {NoStop}%
\bibitem [{\citenamefont {Vidal}\ \emph {et~al.}(2003)\citenamefont {Vidal},
  \citenamefont {Latorre}, \citenamefont {Rico},\ and\ \citenamefont
  {Kitaev}}]{Kitaev2003}%
  \BibitemOpen
  \bibfield  {author} {\bibinfo {author} {\bibfnamefont {G.}~\bibnamefont
  {Vidal}}, \bibinfo {author} {\bibfnamefont {J.~I.}\ \bibnamefont {Latorre}},
  \bibinfo {author} {\bibfnamefont {E.}~\bibnamefont {Rico}}, \ and\ \bibinfo
  {author} {\bibfnamefont {A.}~\bibnamefont {Kitaev}},\ }\bibfield  {title}
  {\enquote {\bibinfo {title} {Entanglement in quantum critical phenomena},}\
  }\href {\doibase 10.1103/PhysRevLett.90.227902} {\bibfield  {journal}
  {\bibinfo  {journal} {Phys. Rev. Lett.}\ }\textbf {\bibinfo {volume} {90}},\
  \bibinfo {pages} {227902} (\bibinfo {year} {2003})}\BibitemShut {NoStop}%
\bibitem [{\citenamefont {Calabrese}\ and\ \citenamefont
  {Cardy}(2004)}]{CalabreseCardy2004}%
  \BibitemOpen
  \bibfield  {author} {\bibinfo {author} {\bibfnamefont {Pasquale}\
  \bibnamefont {Calabrese}}\ and\ \bibinfo {author} {\bibfnamefont {John}\
  \bibnamefont {Cardy}},\ }\bibfield  {title} {\enquote {\bibinfo {title}
  {Entanglement entropy and quantum field theory},}\ }\href {\doibase
  10.1088/1742-5468/2004/06/p06002} {\ \textbf {\bibinfo {volume} {2004}},\
  \bibinfo {pages} {P06002} (\bibinfo {year} {2004})}\BibitemShut {NoStop}%
\bibitem [{\citenamefont {Terhal}\ \emph {et~al.}(2003)\citenamefont {Terhal},
  \citenamefont {Wolf},\ and\ \citenamefont
  {Doherty}}]{terhal2003entanglement}%
  \BibitemOpen
  \bibfield  {author} {\bibinfo {author} {\bibfnamefont {Barbara~M}\
  \bibnamefont {Terhal}}, \bibinfo {author} {\bibfnamefont {Michael~M}\
  \bibnamefont {Wolf}}, \ and\ \bibinfo {author} {\bibfnamefont {Andrew~C}\
  \bibnamefont {Doherty}},\ }\bibfield  {title} {\enquote {\bibinfo {title}
  {Quantum entanglement: A modern perspective},}\ }\href {\doibase
  10.1063/1.1580049} {\bibfield  {journal} {\bibinfo  {journal} {Physics
  Today}\ }\textbf {\bibinfo {volume} {56}},\ \bibinfo {pages} {46--52}
  (\bibinfo {year} {2003})}\BibitemShut {NoStop}%
\bibitem [{\citenamefont {Amico}\ \emph {et~al.}(2008)\citenamefont {Amico},
  \citenamefont {Fazio}, \citenamefont {Osterloh},\ and\ \citenamefont
  {Vedral}}]{vedral2008rmp}%
  \BibitemOpen
  \bibfield  {author} {\bibinfo {author} {\bibfnamefont {Luigi}\ \bibnamefont
  {Amico}}, \bibinfo {author} {\bibfnamefont {Rosario}\ \bibnamefont {Fazio}},
  \bibinfo {author} {\bibfnamefont {Andreas}\ \bibnamefont {Osterloh}}, \ and\
  \bibinfo {author} {\bibfnamefont {Vlatko}\ \bibnamefont {Vedral}},\
  }\bibfield  {title} {\enquote {\bibinfo {title} {Entanglement in many-body
  systems},}\ }\href {\doibase 10.1103/RevModPhys.80.517} {\bibfield  {journal}
  {\bibinfo  {journal} {Rev. Mod. Phys.}\ }\textbf {\bibinfo {volume} {80}},\
  \bibinfo {pages} {517--576} (\bibinfo {year} {2008})}\BibitemShut {NoStop}%
\bibitem [{\citenamefont {Horodecki}\ \emph {et~al.}(2009)\citenamefont
  {Horodecki}, \citenamefont {Horodecki}, \citenamefont {Horodecki},\ and\
  \citenamefont {Horodecki}}]{Horodecki}%
  \BibitemOpen
  \bibfield  {author} {\bibinfo {author} {\bibfnamefont {Ryszard}\ \bibnamefont
  {Horodecki}}, \bibinfo {author} {\bibfnamefont {Pawe\l{}}\ \bibnamefont
  {Horodecki}}, \bibinfo {author} {\bibfnamefont {Micha\l{}}\ \bibnamefont
  {Horodecki}}, \ and\ \bibinfo {author} {\bibfnamefont {Karol}\ \bibnamefont
  {Horodecki}},\ }\bibfield  {title} {\enquote {\bibinfo {title} {Quantum
  entanglement},}\ }\href {\doibase 10.1103/RevModPhys.81.865} {\bibfield
  {journal} {\bibinfo  {journal} {Rev. Mod. Phys.}\ }\textbf {\bibinfo {volume}
  {81}},\ \bibinfo {pages} {865} (\bibinfo {year} {2009})}\BibitemShut
  {NoStop}%
\bibitem [{\citenamefont {Eisert}\ \emph {et~al.}(2010)\citenamefont {Eisert},
  \citenamefont {Cramer},\ and\ \citenamefont {Plenio}}]{Eisert2010}%
  \BibitemOpen
  \bibfield  {author} {\bibinfo {author} {\bibfnamefont {J.}~\bibnamefont
  {Eisert}}, \bibinfo {author} {\bibfnamefont {M.}~\bibnamefont {Cramer}}, \
  and\ \bibinfo {author} {\bibfnamefont {M.~B.}\ \bibnamefont {Plenio}},\
  }\bibfield  {title} {\enquote {\bibinfo {title} {Colloquium: Area laws for
  the entanglement entropy},}\ }\href {\doibase 10.1103/RevModPhys.82.277}
  {\bibfield  {journal} {\bibinfo  {journal} {Rev. Mod. Phys.}\ }\textbf
  {\bibinfo {volume} {82}},\ \bibinfo {pages} {277} (\bibinfo {year}
  {2010})}\BibitemShut {NoStop}%
\bibitem [{\citenamefont {Hastings}(2007)}]{hastings2007area}%
  \BibitemOpen
  \bibfield  {author} {\bibinfo {author} {\bibfnamefont {Matthew~B}\
  \bibnamefont {Hastings}},\ }\bibfield  {title} {\enquote {\bibinfo {title}
  {An area law for one-dimensional quantum systems},}\ }\href {\doibase
  10.1088/1742-5468/2007/08/P08024} {\bibfield  {journal} {\bibinfo  {journal}
  {J. Stat. Mech.: Theory Exp.}\ }\textbf {\bibinfo {volume} {2007}},\ \bibinfo
  {pages} {P08024} (\bibinfo {year} {2007})}\BibitemShut {NoStop}%
\bibitem [{\citenamefont {Plenio}\ \emph {et~al.}(2005)\citenamefont {Plenio},
  \citenamefont {Eisert}, \citenamefont {Drei\ss{}ig},\ and\ \citenamefont
  {Cramer}}]{Plenio2005}%
  \BibitemOpen
  \bibfield  {author} {\bibinfo {author} {\bibfnamefont {M.~B.}\ \bibnamefont
  {Plenio}}, \bibinfo {author} {\bibfnamefont {J.}~\bibnamefont {Eisert}},
  \bibinfo {author} {\bibfnamefont {J.}~\bibnamefont {Drei\ss{}ig}}, \ and\
  \bibinfo {author} {\bibfnamefont {M.}~\bibnamefont {Cramer}},\ }\bibfield
  {title} {\enquote {\bibinfo {title} {Entropy, entanglement, and area:
  Analytical results for harmonic lattice systems},}\ }\href {\doibase
  10.1103/PhysRevLett.94.060503} {\bibfield  {journal} {\bibinfo  {journal}
  {Phys. Rev. Lett.}\ }\textbf {\bibinfo {volume} {94}},\ \bibinfo {pages}
  {060503} (\bibinfo {year} {2005})}\BibitemShut {NoStop}%
\bibitem [{\citenamefont {Bisognano}\ and\ \citenamefont
  {Wichmann}(1975)}]{BW1975}%
  \BibitemOpen
  \bibfield  {author} {\bibinfo {author} {\bibfnamefont {Joseph~J}\
  \bibnamefont {Bisognano}}\ and\ \bibinfo {author} {\bibfnamefont {Eyvind~H}\
  \bibnamefont {Wichmann}},\ }\bibfield  {title} {\enquote {\bibinfo {title}
  {On the duality condition for a hermitian scalar field},}\ }\href@noop {}
  {\bibfield  {journal} {\bibinfo  {journal} {Journal of Mathematical Physics}\
  }\textbf {\bibinfo {volume} {16}},\ \bibinfo {pages} {985} (\bibinfo {year}
  {1975})}\BibitemShut {NoStop}%
\bibitem [{\citenamefont {Swingle}\ and\ \citenamefont
  {McGreevy}(2016)}]{Swingle2016}%
  \BibitemOpen
  \bibfield  {author} {\bibinfo {author} {\bibfnamefont {Brian}\ \bibnamefont
  {Swingle}}\ and\ \bibinfo {author} {\bibfnamefont {John}\ \bibnamefont
  {McGreevy}},\ }\bibfield  {title} {\enquote {\bibinfo {title} {Area law for
  gapless states from local entanglement thermodynamics},}\ }\href {\doibase
  10.1103/PhysRevB.93.205120} {\bibfield  {journal} {\bibinfo  {journal} {Phys.
  Rev. B}\ }\textbf {\bibinfo {volume} {93}},\ \bibinfo {pages} {205120}
  (\bibinfo {year} {2016})}\BibitemShut {NoStop}%
\bibitem [{\citenamefont {Dalmonte}\ \emph {et~al.}(2018)\citenamefont
  {Dalmonte}, \citenamefont {Vermersch},\ and\ \citenamefont
  {Zoller}}]{dalmonte2018quantum}%
  \BibitemOpen
  \bibfield  {author} {\bibinfo {author} {\bibfnamefont {Marcello}\
  \bibnamefont {Dalmonte}}, \bibinfo {author} {\bibfnamefont {Beno{\^\i}t}\
  \bibnamefont {Vermersch}}, \ and\ \bibinfo {author} {\bibfnamefont {Peter}\
  \bibnamefont {Zoller}},\ }\bibfield  {title} {\enquote {\bibinfo {title}
  {Quantum simulation and spectroscopy of entanglement hamiltonians},}\ }\href
  {\doibase 10.1038/s41567-018-0151-7} {\bibfield  {journal} {\bibinfo
  {journal} {Nat. Phys.}\ }\textbf {\bibinfo {volume} {14}},\ \bibinfo {pages}
  {827} (\bibinfo {year} {2018})}\BibitemShut {NoStop}%
\bibitem [{\citenamefont {Dalmonte}\ \emph {et~al.}(2022)\citenamefont
  {Dalmonte}, \citenamefont {Eisler}, \citenamefont {Falconi},\ and\
  \citenamefont {Vermersch}}]{dalmoste2022}%
  \BibitemOpen
  \bibfield  {author} {\bibinfo {author} {\bibfnamefont {M.}~\bibnamefont
  {Dalmonte}}, \bibinfo {author} {\bibfnamefont {V.}~\bibnamefont {Eisler}},
  \bibinfo {author} {\bibfnamefont {M.}~\bibnamefont {Falconi}}, \ and\
  \bibinfo {author} {\bibfnamefont {B.}~\bibnamefont {Vermersch}},\ }\href
  {https://doi.org/10.48550/arXiv.2202.05045} {\enquote {\bibinfo {title}
  {Entanglement hamiltonians: from field theory, to lattice models and
  experiments},}\ } (\bibinfo {year} {2022}),\ \Eprint
  {http://arxiv.org/abs/2202.05045} {arXiv:2202.05045 [cond-mat.stat-mech]}
  \BibitemShut {NoStop}%
\bibitem [{\citenamefont {Pourjafarabadi}\ \emph {et~al.}(2021)\citenamefont
  {Pourjafarabadi}, \citenamefont {Najafzadeh}, \citenamefont {Vaezi},\ and\
  \citenamefont {Vaezi}}]{vaezi2021}%
  \BibitemOpen
  \bibfield  {author} {\bibinfo {author} {\bibfnamefont {Mahdieh}\ \bibnamefont
  {Pourjafarabadi}}, \bibinfo {author} {\bibfnamefont {Hanieh}\ \bibnamefont
  {Najafzadeh}}, \bibinfo {author} {\bibfnamefont {Mohammad-Sadegh}\
  \bibnamefont {Vaezi}}, \ and\ \bibinfo {author} {\bibfnamefont {Abolhassan}\
  \bibnamefont {Vaezi}},\ }\bibfield  {title} {\enquote {\bibinfo {title}
  {Entanglement hamiltonian of interacting systems: Local temperature
  approximation and beyond},}\ }\href {\doibase
  10.1103/PhysRevResearch.3.013217} {\bibfield  {journal} {\bibinfo  {journal}
  {Phys. Rev. Research}\ }\textbf {\bibinfo {volume} {3}},\ \bibinfo {pages}
  {013217} (\bibinfo {year} {2021})}\BibitemShut {NoStop}%
\bibitem [{\citenamefont {Rigol}\ \emph {et~al.}(2007)\citenamefont {Rigol},
  \citenamefont {Dunjko}, \citenamefont {Yurovsky},\ and\ \citenamefont
  {Olshanii}}]{Olshanii2007}%
  \BibitemOpen
  \bibfield  {author} {\bibinfo {author} {\bibfnamefont {Marcos}\ \bibnamefont
  {Rigol}}, \bibinfo {author} {\bibfnamefont {Vanja}\ \bibnamefont {Dunjko}},
  \bibinfo {author} {\bibfnamefont {Vladimir}\ \bibnamefont {Yurovsky}}, \ and\
  \bibinfo {author} {\bibfnamefont {Maxim}\ \bibnamefont {Olshanii}},\
  }\bibfield  {title} {\enquote {\bibinfo {title} {Relaxation in a completely
  integrable many-body quantum system: An ab initio study of the dynamics of
  the highly excited states of 1d lattice hard-core bosons},}\ }\href {\doibase
  10.1103/PhysRevLett.98.050405} {\bibfield  {journal} {\bibinfo  {journal}
  {Phys. Rev. Lett.}\ }\textbf {\bibinfo {volume} {98}},\ \bibinfo {pages}
  {050405} (\bibinfo {year} {2007})}\BibitemShut {NoStop}%
\bibitem [{\citenamefont {Peschel}(2003)}]{Peschel_2003}%
  \BibitemOpen
  \bibfield  {author} {\bibinfo {author} {\bibfnamefont {Ingo}\ \bibnamefont
  {Peschel}},\ }\bibfield  {title} {\enquote {\bibinfo {title} {Calculation of
  reduced density matrices from correlation functions},}\ }\href {\doibase
  10.1088/0305-4470/36/14/101} {\bibfield  {journal} {\bibinfo  {journal} {J.
  Phys. A}\ }\textbf {\bibinfo {volume} {36}},\ \bibinfo {pages} {L205}
  (\bibinfo {year} {2003})}\BibitemShut {NoStop}%
\bibitem [{\citenamefont {Peschel}\ and\ \citenamefont
  {Eisler}(2009)}]{Peschel_2009}%
  \BibitemOpen
  \bibfield  {author} {\bibinfo {author} {\bibfnamefont {Ingo}\ \bibnamefont
  {Peschel}}\ and\ \bibinfo {author} {\bibfnamefont {Viktor}\ \bibnamefont
  {Eisler}},\ }\bibfield  {title} {\enquote {\bibinfo {title} {Reduced density
  matrices and entanglement entropy in free lattice models},}\ }\href {\doibase
  10.1088/1751-8113/42/50/504003} {\bibfield  {journal} {\bibinfo  {journal}
  {Journal of Physics A: Mathematical and Theoretical}\ }\textbf {\bibinfo
  {volume} {42}},\ \bibinfo {pages} {504003} (\bibinfo {year}
  {2009})}\BibitemShut {NoStop}%
\bibitem [{\citenamefont {Schnyder}\ \emph {et~al.}(2008)\citenamefont
  {Schnyder}, \citenamefont {Ryu}, \citenamefont {Furusaki},\ and\
  \citenamefont {Ludwig}}]{Schnyder2008}%
  \BibitemOpen
  \bibfield  {author} {\bibinfo {author} {\bibfnamefont {Andreas~P.}\
  \bibnamefont {Schnyder}}, \bibinfo {author} {\bibfnamefont {Shinsei}\
  \bibnamefont {Ryu}}, \bibinfo {author} {\bibfnamefont {Akira}\ \bibnamefont
  {Furusaki}}, \ and\ \bibinfo {author} {\bibfnamefont {Andreas W.~W.}\
  \bibnamefont {Ludwig}},\ }\bibfield  {title} {\enquote {\bibinfo {title}
  {Classification of topological insulators and superconductors in three
  spatial dimensions},}\ }\href {\doibase 10.1103/PhysRevB.78.195125}
  {\bibfield  {journal} {\bibinfo  {journal} {Phys. Rev. B}\ }\textbf {\bibinfo
  {volume} {78}},\ \bibinfo {pages} {195125} (\bibinfo {year}
  {2008})}\BibitemShut {NoStop}%
\bibitem [{\citenamefont {Kitaev}(2009)}]{kitaev2009}%
  \BibitemOpen
  \bibfield  {author} {\bibinfo {author} {\bibfnamefont {Alexei}\ \bibnamefont
  {Kitaev}},\ }\bibfield  {title} {\enquote {\bibinfo {title} {Periodic table
  for topological insulators and superconductors},}\ }\href {\doibase
  10.1063/1.3149495} {\bibfield  {journal} {\bibinfo  {journal} {AIP Conference
  Proceedings}\ }\textbf {\bibinfo {volume} {1134}},\ \bibinfo {pages} {22--30}
  (\bibinfo {year} {2009})}\BibitemShut {NoStop}%
\bibitem [{\citenamefont {Fulga}\ \emph {et~al.}(2012)\citenamefont {Fulga},
  \citenamefont {Hassler},\ and\ \citenamefont {Akhmerov}}]{fulga2012}%
  \BibitemOpen
  \bibfield  {author} {\bibinfo {author} {\bibfnamefont {I.~C.}\ \bibnamefont
  {Fulga}}, \bibinfo {author} {\bibfnamefont {F.}~\bibnamefont {Hassler}}, \
  and\ \bibinfo {author} {\bibfnamefont {A.~R.}\ \bibnamefont {Akhmerov}},\
  }\bibfield  {title} {\enquote {\bibinfo {title} {Scattering theory of
  topological insulators and superconductors},}\ }\href {\doibase
  10.1103/PhysRevB.85.165409} {\bibfield  {journal} {\bibinfo  {journal} {Phys.
  Rev. B}\ }\textbf {\bibinfo {volume} {85}},\ \bibinfo {pages} {165409}
  (\bibinfo {year} {2012})}\BibitemShut {NoStop}%
\bibitem [{\citenamefont {Zilberberg}\ \emph {et~al.}(2018)\citenamefont
  {Zilberberg}, \citenamefont {Huang}, \citenamefont {Guglielmon},
  \citenamefont {Wang}, \citenamefont {Chen}, \citenamefont {Kraus},\ and\
  \citenamefont {Rechtsman}}]{Zilberberg2018}%
  \BibitemOpen
  \bibfield  {author} {\bibinfo {author} {\bibfnamefont {Oded}\ \bibnamefont
  {Zilberberg}}, \bibinfo {author} {\bibfnamefont {Sheng}\ \bibnamefont
  {Huang}}, \bibinfo {author} {\bibfnamefont {Jonathan}\ \bibnamefont
  {Guglielmon}}, \bibinfo {author} {\bibfnamefont {Mohan}\ \bibnamefont
  {Wang}}, \bibinfo {author} {\bibfnamefont {Kevin~P.}\ \bibnamefont {Chen}},
  \bibinfo {author} {\bibfnamefont {Yaacov~E.}\ \bibnamefont {Kraus}}, \ and\
  \bibinfo {author} {\bibfnamefont {Mikael~C.}\ \bibnamefont {Rechtsman}},\
  }\bibfield  {title} {\enquote {\bibinfo {title} {Photonic topological
  boundary pumping as a probe of 4d quantum hall physics},}\ }\href {\doibase
  10.1038/nature25011} {\bibfield  {journal} {\bibinfo  {journal} {Nature}\
  }\textbf {\bibinfo {volume} {553}},\ \bibinfo {pages} {59--62} (\bibinfo
  {year} {2018})}\BibitemShut {NoStop}%
\bibitem [{\citenamefont {Lohse}\ \emph {et~al.}(2018)\citenamefont {Lohse},
  \citenamefont {Schweizer}, \citenamefont {Price}, \citenamefont
  {Zilberberg},\ and\ \citenamefont {Bloch}}]{Lohse2018}%
  \BibitemOpen
  \bibfield  {author} {\bibinfo {author} {\bibfnamefont {Michael}\ \bibnamefont
  {Lohse}}, \bibinfo {author} {\bibfnamefont {Christian}\ \bibnamefont
  {Schweizer}}, \bibinfo {author} {\bibfnamefont {Hannah~M.}\ \bibnamefont
  {Price}}, \bibinfo {author} {\bibfnamefont {Oded}\ \bibnamefont
  {Zilberberg}}, \ and\ \bibinfo {author} {\bibfnamefont {Immanuel}\
  \bibnamefont {Bloch}},\ }\bibfield  {title} {\enquote {\bibinfo {title}
  {Exploring 4d quantum hall physics with a 2d topological charge pump},}\
  }\href {\doibase 10.1038/nature25000} {\bibfield  {journal} {\bibinfo
  {journal} {Nature}\ }\textbf {\bibinfo {volume} {553}},\ \bibinfo {pages}
  {55--58} (\bibinfo {year} {2018})}\BibitemShut {NoStop}%
\bibitem [{\citenamefont {Zhang}\ and\ \citenamefont {Hu}(2001)}]{zhang2001}%
  \BibitemOpen
  \bibfield  {author} {\bibinfo {author} {\bibfnamefont {Shou-Cheng}\
  \bibnamefont {Zhang}}\ and\ \bibinfo {author} {\bibfnamefont {Jiangping}\
  \bibnamefont {Hu}},\ }\bibfield  {title} {\enquote {\bibinfo {title} {A
  four-dimensional generalization of the quantum hall effect},}\ }\href
  {\doibase 10.1126/science.294.5543.823} {\bibfield  {journal} {\bibinfo
  {journal} {Science}\ }\textbf {\bibinfo {volume} {294}},\ \bibinfo {pages}
  {823} (\bibinfo {year} {2001})}\BibitemShut {NoStop}%
\bibitem [{\citenamefont {Qi}\ \emph {et~al.}(2008)\citenamefont {Qi},
  \citenamefont {Hughes},\ and\ \citenamefont {Zhang}}]{qi-zhang2008}%
  \BibitemOpen
  \bibfield  {author} {\bibinfo {author} {\bibfnamefont {Xiao-Liang}\
  \bibnamefont {Qi}}, \bibinfo {author} {\bibfnamefont {Taylor~L.}\
  \bibnamefont {Hughes}}, \ and\ \bibinfo {author} {\bibfnamefont {Shou-Cheng}\
  \bibnamefont {Zhang}},\ }\bibfield  {title} {\enquote {\bibinfo {title}
  {Topological field theory of time-reversal invariant insulators},}\ }\href
  {\doibase 10.1103/PhysRevB.78.195424} {\bibfield  {journal} {\bibinfo
  {journal} {Phys. Rev. B}\ }\textbf {\bibinfo {volume} {78}},\ \bibinfo
  {pages} {195424} (\bibinfo {year} {2008})}\BibitemShut {NoStop}%
\bibitem [{\citenamefont {Qi}\ and\ \citenamefont
  {Zhang}(2011)}]{qi-zhang-2011}%
  \BibitemOpen
  \bibfield  {author} {\bibinfo {author} {\bibfnamefont {Xiao-Liang}\
  \bibnamefont {Qi}}\ and\ \bibinfo {author} {\bibfnamefont {Shou-Cheng}\
  \bibnamefont {Zhang}},\ }\bibfield  {title} {\enquote {\bibinfo {title}
  {Topological insulators and superconductors},}\ }\href {\doibase
  10.1103/RevModPhys.83.1057} {\bibfield  {journal} {\bibinfo  {journal} {Rev.
  Mod. Phys.}\ }\textbf {\bibinfo {volume} {83}},\ \bibinfo {pages}
  {1057--1110} (\bibinfo {year} {2011})}\BibitemShut {NoStop}%
\bibitem [{\citenamefont {Chiu}\ \emph {et~al.}(2016)\citenamefont {Chiu},
  \citenamefont {Teo}, \citenamefont {Schnyder},\ and\ \citenamefont
  {Ryu}}]{Ryu2016}%
  \BibitemOpen
  \bibfield  {author} {\bibinfo {author} {\bibfnamefont {Ching-Kai}\
  \bibnamefont {Chiu}}, \bibinfo {author} {\bibfnamefont {Jeffrey C.~Y.}\
  \bibnamefont {Teo}}, \bibinfo {author} {\bibfnamefont {Andreas~P.}\
  \bibnamefont {Schnyder}}, \ and\ \bibinfo {author} {\bibfnamefont {Shinsei}\
  \bibnamefont {Ryu}},\ }\bibfield  {title} {\enquote {\bibinfo {title}
  {Classification of topological quantum matter with symmetries},}\ }\href
  {\doibase 10.1103/RevModPhys.88.035005} {\bibfield  {journal} {\bibinfo
  {journal} {Rev. Mod. Phys.}\ }\textbf {\bibinfo {volume} {88}},\ \bibinfo
  {pages} {035005} (\bibinfo {year} {2016})}\BibitemShut {NoStop}%
\bibitem [{foo()}]{footnote1}%
  \BibitemOpen
  \bibinfo {note} {The invariant given in terms of the map $\hat{\bf d}_{\bf
  k}$ has a simple geometric meaning as the number of times the unit vector
  $\hat{\bf d}_{\bf k}$ wraps around the $n$-sphere when ${\bf k}$ sweeps the
  whole Brillouin zone. Such a intuitive interpretation of the topological
  invariant translates to the more rigorous understanding that the $n^{\rm th}$
  homotopy class of $n$-sphere is equivalent to ${\mathbb Z}$:
  $\pi_{n}(S^n)\equiv {\mathbb Z}$.}\BibitemShut {Stop}%
\bibitem [{\citenamefont {Kaufman}\ \emph {et~al.}(2016)\citenamefont
  {Kaufman}, \citenamefont {Tai}, \citenamefont {Lukin}, \citenamefont
  {Rispoli}, \citenamefont {Schittko}, \citenamefont {Preiss},\ and\
  \citenamefont {Greiner}}]{Greiner2016quantum}%
  \BibitemOpen
  \bibfield  {author} {\bibinfo {author} {\bibfnamefont {Adam~M}\ \bibnamefont
  {Kaufman}}, \bibinfo {author} {\bibfnamefont {M~Eric}\ \bibnamefont {Tai}},
  \bibinfo {author} {\bibfnamefont {Alexander}\ \bibnamefont {Lukin}}, \bibinfo
  {author} {\bibfnamefont {Matthew}\ \bibnamefont {Rispoli}}, \bibinfo {author}
  {\bibfnamefont {Robert}\ \bibnamefont {Schittko}}, \bibinfo {author}
  {\bibfnamefont {Philipp~M}\ \bibnamefont {Preiss}}, \ and\ \bibinfo {author}
  {\bibfnamefont {Markus}\ \bibnamefont {Greiner}},\ }\bibfield  {title}
  {\enquote {\bibinfo {title} {Quantum thermalization through entanglement in
  an isolated many-body system},}\ }\href {\doibase 10.1126/science.aaf6725}
  {\bibfield  {journal} {\bibinfo  {journal} {Science}\ }\textbf {\bibinfo
  {volume} {353}},\ \bibinfo {pages} {794} (\bibinfo {year}
  {2016})}\BibitemShut {NoStop}%
\bibitem [{\citenamefont {Islam}\ \emph {et~al.}(2015)\citenamefont {Islam},
  \citenamefont {Ma}, \citenamefont {Preiss}, \citenamefont {Eric~Tai},
  \citenamefont {Lukin}, \citenamefont {Rispoli},\ and\ \citenamefont
  {Greiner}}]{Greiner2015measuring}%
  \BibitemOpen
  \bibfield  {author} {\bibinfo {author} {\bibfnamefont {Rajibul}\ \bibnamefont
  {Islam}}, \bibinfo {author} {\bibfnamefont {Ruichao}\ \bibnamefont {Ma}},
  \bibinfo {author} {\bibfnamefont {Philipp~M}\ \bibnamefont {Preiss}},
  \bibinfo {author} {\bibfnamefont {M}~\bibnamefont {Eric~Tai}}, \bibinfo
  {author} {\bibfnamefont {Alexander}\ \bibnamefont {Lukin}}, \bibinfo {author}
  {\bibfnamefont {Matthew}\ \bibnamefont {Rispoli}}, \ and\ \bibinfo {author}
  {\bibfnamefont {Markus}\ \bibnamefont {Greiner}},\ }\bibfield  {title}
  {\enquote {\bibinfo {title} {Measuring entanglement entropy in a quantum
  many-body system},}\ }\href {\doibase 10.1038/nature15750} {\bibfield
  {journal} {\bibinfo  {journal} {Nature}\ }\textbf {\bibinfo {volume} {528}},\
  \bibinfo {pages} {77} (\bibinfo {year} {2015})}\BibitemShut {NoStop}%
\bibitem [{\citenamefont {Gross}\ and\ \citenamefont
  {Bloch}(2017)}]{Bloch2017quantum}%
  \BibitemOpen
  \bibfield  {author} {\bibinfo {author} {\bibfnamefont {Christian}\
  \bibnamefont {Gross}}\ and\ \bibinfo {author} {\bibfnamefont {Immanuel}\
  \bibnamefont {Bloch}},\ }\bibfield  {title} {\enquote {\bibinfo {title}
  {Quantum simulations with ultracold atoms in optical lattices},}\ }\href
  {\doibase 10.1126/science.aal3837} {\bibfield  {journal} {\bibinfo  {journal}
  {Science}\ }\textbf {\bibinfo {volume} {357}},\ \bibinfo {pages} {995}
  (\bibinfo {year} {2017})}\BibitemShut {NoStop}%
\bibitem [{\citenamefont {Brydges}\ \emph {et~al.}(2019)\citenamefont
  {Brydges}, \citenamefont {Elben}, \citenamefont {Jurcevic}, \citenamefont
  {Vermersch}, \citenamefont {Maier}, \citenamefont {Lanyon}, \citenamefont
  {Zoller}, \citenamefont {Blatt},\ and\ \citenamefont
  {Roos}}]{brydges2019probing}%
  \BibitemOpen
  \bibfield  {author} {\bibinfo {author} {\bibfnamefont {Tiff}\ \bibnamefont
  {Brydges}}, \bibinfo {author} {\bibfnamefont {Andreas}\ \bibnamefont
  {Elben}}, \bibinfo {author} {\bibfnamefont {Petar}\ \bibnamefont {Jurcevic}},
  \bibinfo {author} {\bibfnamefont {Beno{\^\i}t}\ \bibnamefont {Vermersch}},
  \bibinfo {author} {\bibfnamefont {Christine}\ \bibnamefont {Maier}}, \bibinfo
  {author} {\bibfnamefont {Ben~P}\ \bibnamefont {Lanyon}}, \bibinfo {author}
  {\bibfnamefont {Peter}\ \bibnamefont {Zoller}}, \bibinfo {author}
  {\bibfnamefont {Rainer}\ \bibnamefont {Blatt}}, \ and\ \bibinfo {author}
  {\bibfnamefont {Christian~F}\ \bibnamefont {Roos}},\ }\bibfield  {title}
  {\enquote {\bibinfo {title} {Probing r{\'e}nyi entanglement entropy via
  randomized measurements},}\ }\href {\doibase 10.1126/science.aau4963}
  {\bibfield  {journal} {\bibinfo  {journal} {Science}\ }\textbf {\bibinfo
  {volume} {364}},\ \bibinfo {pages} {260} (\bibinfo {year}
  {2019})}\BibitemShut {NoStop}%
\bibitem [{\citenamefont {Jotzu}\ \emph {et~al.}(2014)\citenamefont {Jotzu},
  \citenamefont {Messer}, \citenamefont {Desbuquois}, \citenamefont {Lebrat},
  \citenamefont {Uehlinger}, \citenamefont {Greif},\ and\ \citenamefont
  {Esslinger}}]{jotzu2014Haldane}%
  \BibitemOpen
  \bibfield  {author} {\bibinfo {author} {\bibfnamefont {Gregor}\ \bibnamefont
  {Jotzu}}, \bibinfo {author} {\bibfnamefont {Michael}\ \bibnamefont {Messer}},
  \bibinfo {author} {\bibfnamefont {R{\'e}mi}\ \bibnamefont {Desbuquois}},
  \bibinfo {author} {\bibfnamefont {Martin}\ \bibnamefont {Lebrat}}, \bibinfo
  {author} {\bibfnamefont {Thomas}\ \bibnamefont {Uehlinger}}, \bibinfo
  {author} {\bibfnamefont {Daniel}\ \bibnamefont {Greif}}, \ and\ \bibinfo
  {author} {\bibfnamefont {Tilman}\ \bibnamefont {Esslinger}},\ }\bibfield
  {title} {\enquote {\bibinfo {title} {Experimental realization of the
  topological haldane model with ultracold fermions},}\ }\href {\doibase
  10.1038/nature13915} {\bibfield  {journal} {\bibinfo  {journal} {Nature}\
  }\textbf {\bibinfo {volume} {515}},\ \bibinfo {pages} {237--240} (\bibinfo
  {year} {2014})}\BibitemShut {NoStop}%
\bibitem [{\citenamefont {Aidelsburger}\ \emph {et~al.}(2015)\citenamefont
  {Aidelsburger}, \citenamefont {Lohse}, \citenamefont {Schweizer},
  \citenamefont {Atala}, \citenamefont {Barreiro}, \citenamefont
  {Nascimb{\`e}ne}, \citenamefont {Cooper}, \citenamefont {Bloch},\ and\
  \citenamefont {Goldman}}]{aidelsburger2015measuring}%
  \BibitemOpen
  \bibfield  {author} {\bibinfo {author} {\bibfnamefont {Monika}\ \bibnamefont
  {Aidelsburger}}, \bibinfo {author} {\bibfnamefont {Michael}\ \bibnamefont
  {Lohse}}, \bibinfo {author} {\bibfnamefont {Christian}\ \bibnamefont
  {Schweizer}}, \bibinfo {author} {\bibfnamefont {Marcos}\ \bibnamefont
  {Atala}}, \bibinfo {author} {\bibfnamefont {Julio~T}\ \bibnamefont
  {Barreiro}}, \bibinfo {author} {\bibfnamefont {Sylvain}\ \bibnamefont
  {Nascimb{\`e}ne}}, \bibinfo {author} {\bibfnamefont {NR}~\bibnamefont
  {Cooper}}, \bibinfo {author} {\bibfnamefont {Immanuel}\ \bibnamefont
  {Bloch}}, \ and\ \bibinfo {author} {\bibfnamefont {Nathan}\ \bibnamefont
  {Goldman}},\ }\bibfield  {title} {\enquote {\bibinfo {title} {Measuring the
  chern number of hofstadter bands with ultracold bosonic atoms},}\ }\href
  {\doibase 10.1038/nphys3171} {\bibfield  {journal} {\bibinfo  {journal}
  {Nature Physics}\ }\textbf {\bibinfo {volume} {11}},\ \bibinfo {pages}
  {162--166} (\bibinfo {year} {2015})}\BibitemShut {NoStop}%
\bibitem [{\citenamefont {Goldman}\ \emph {et~al.}(2016)\citenamefont
  {Goldman}, \citenamefont {Budich},\ and\ \citenamefont
  {Zoller}}]{goldman2016topological}%
  \BibitemOpen
  \bibfield  {author} {\bibinfo {author} {\bibfnamefont {Nathan}\ \bibnamefont
  {Goldman}}, \bibinfo {author} {\bibfnamefont {Jan~C}\ \bibnamefont {Budich}},
  \ and\ \bibinfo {author} {\bibfnamefont {Peter}\ \bibnamefont {Zoller}},\
  }\bibfield  {title} {\enquote {\bibinfo {title} {Topological quantum matter
  with ultracold gases in optical lattices},}\ }\href {\doibase
  10.1038/nphys3803} {\bibfield  {journal} {\bibinfo  {journal} {Nature
  Physics}\ }\textbf {\bibinfo {volume} {12}},\ \bibinfo {pages} {639--645}
  (\bibinfo {year} {2016})}\BibitemShut {NoStop}%
\bibitem [{\citenamefont {Wu}\ \emph {et~al.}(2016)\citenamefont {Wu},
  \citenamefont {Zhang}, \citenamefont {Sun}, \citenamefont {Xu}, \citenamefont
  {Wang}, \citenamefont {Ji}, \citenamefont {Deng}, \citenamefont {Chen},
  \citenamefont {Liu},\ and\ \citenamefont {Pan}}]{wu2016realization}%
  \BibitemOpen
  \bibfield  {author} {\bibinfo {author} {\bibfnamefont {Zhan}\ \bibnamefont
  {Wu}}, \bibinfo {author} {\bibfnamefont {Long}\ \bibnamefont {Zhang}},
  \bibinfo {author} {\bibfnamefont {Wei}\ \bibnamefont {Sun}}, \bibinfo
  {author} {\bibfnamefont {Xiao-Tian}\ \bibnamefont {Xu}}, \bibinfo {author}
  {\bibfnamefont {Bao-Zong}\ \bibnamefont {Wang}}, \bibinfo {author}
  {\bibfnamefont {Si-Cong}\ \bibnamefont {Ji}}, \bibinfo {author}
  {\bibfnamefont {Youjin}\ \bibnamefont {Deng}}, \bibinfo {author}
  {\bibfnamefont {Shuai}\ \bibnamefont {Chen}}, \bibinfo {author}
  {\bibfnamefont {Xiong-Jun}\ \bibnamefont {Liu}}, \ and\ \bibinfo {author}
  {\bibfnamefont {Jian-Wei}\ \bibnamefont {Pan}},\ }\bibfield  {title}
  {\enquote {\bibinfo {title} {Realization of two-dimensional spin-orbit
  coupling for bose-einstein condensates},}\ }\href {\doibase
  10.1126/science.aaf6689} {\bibfield  {journal} {\bibinfo  {journal}
  {Science}\ }\textbf {\bibinfo {volume} {354}},\ \bibinfo {pages} {83--88}
  (\bibinfo {year} {2016})}\BibitemShut {NoStop}%
\bibitem [{\citenamefont {Cooper}\ \emph {et~al.}(2019)\citenamefont {Cooper},
  \citenamefont {Dalibard},\ and\ \citenamefont {Spielman}}]{Cooper2019RMP}%
  \BibitemOpen
  \bibfield  {author} {\bibinfo {author} {\bibfnamefont {N.~R.}\ \bibnamefont
  {Cooper}}, \bibinfo {author} {\bibfnamefont {J.}~\bibnamefont {Dalibard}}, \
  and\ \bibinfo {author} {\bibfnamefont {I.~B.}\ \bibnamefont {Spielman}},\
  }\bibfield  {title} {\enquote {\bibinfo {title} {Topological bands for
  ultracold atoms},}\ }\href {\doibase 10.1103/RevModPhys.91.015005} {\bibfield
   {journal} {\bibinfo  {journal} {Rev. Mod. Phys.}\ }\textbf {\bibinfo
  {volume} {91}},\ \bibinfo {pages} {015005} (\bibinfo {year}
  {2019})}\BibitemShut {NoStop}%
\bibitem [{\citenamefont {Eisler}\ \emph {et~al.}(2006)\citenamefont {Eisler},
  \citenamefont {Legeza},\ and\ \citenamefont
  {R{\'{a}}cz}}]{Eisler2006Fluctuations}%
  \BibitemOpen
  \bibfield  {author} {\bibinfo {author} {\bibfnamefont {V.}~\bibnamefont
  {Eisler}}, \bibinfo {author} {\bibfnamefont {\"O.}\ \bibnamefont {Legeza}}, \
  and\ \bibinfo {author} {\bibfnamefont {Z.}~\bibnamefont {R{\'{a}}cz}},\
  }\bibfield  {title} {\enquote {\bibinfo {title} {{Fluctuations in subsystems
  of the zero-temperature XX chain: emergence of an effective temperature}},}\
  }\href {\doibase 10.1088/1742-5468/2006/11/p11013} {\bibfield  {journal}
  {\bibinfo  {journal} {J. Stat. Mech.: Theory Exp.}\ }\textbf {\bibinfo
  {volume} {2006}},\ \bibinfo {pages} {P11013} (\bibinfo {year}
  {2006})}\BibitemShut {NoStop}%
\bibitem [{Note1()}]{Note1}%
  \BibitemOpen
  \bibinfo {note} {This result reflects another problem with defining
  entanglement temperature for subsystems with the same dimensionality as the
  system, because the resulting average temperature vanishes for very large
  sizes.}\BibitemShut {Stop}%
\bibitem [{\citenamefont {Klich}\ and\ \citenamefont
  {Levitov}(2009)}]{KlichLevitov2009}%
  \BibitemOpen
  \bibfield  {author} {\bibinfo {author} {\bibfnamefont {Israel}\ \bibnamefont
  {Klich}}\ and\ \bibinfo {author} {\bibfnamefont {Leonid}\ \bibnamefont
  {Levitov}},\ }\bibfield  {title} {\enquote {\bibinfo {title} {Quantum noise
  as an entanglement meter},}\ }\href {\doibase 10.1103/PhysRevLett.102.100502}
  {\bibfield  {journal} {\bibinfo  {journal} {Phys. Rev. Lett.}\ }\textbf
  {\bibinfo {volume} {102}},\ \bibinfo {pages} {100502} (\bibinfo {year}
  {2009})}\BibitemShut {NoStop}%
\bibitem [{\citenamefont {Song}\ \emph {et~al.}(2010)\citenamefont {Song},
  \citenamefont {Rachel},\ and\ \citenamefont {Le~Hur}}]{LeHur2010}%
  \BibitemOpen
  \bibfield  {author} {\bibinfo {author} {\bibfnamefont {H.~Francis}\
  \bibnamefont {Song}}, \bibinfo {author} {\bibfnamefont {Stephan}\
  \bibnamefont {Rachel}}, \ and\ \bibinfo {author} {\bibfnamefont {Karyn}\
  \bibnamefont {Le~Hur}},\ }\bibfield  {title} {\enquote {\bibinfo {title}
  {General relation between entanglement and fluctuations in one dimension},}\
  }\href {\doibase 10.1103/PhysRevB.82.012405} {\bibfield  {journal} {\bibinfo
  {journal} {Phys. Rev. B}\ }\textbf {\bibinfo {volume} {82}},\ \bibinfo
  {pages} {012405} (\bibinfo {year} {2010})}\BibitemShut {NoStop}%
\bibitem [{\citenamefont {Song}\ \emph {et~al.}(2012)\citenamefont {Song},
  \citenamefont {Rachel}, \citenamefont {Flindt}, \citenamefont {Klich},
  \citenamefont {Laflorencie},\ and\ \citenamefont {Le~Hur}}]{LeHur2012}%
  \BibitemOpen
  \bibfield  {author} {\bibinfo {author} {\bibfnamefont {H.~Francis}\
  \bibnamefont {Song}}, \bibinfo {author} {\bibfnamefont {Stephan}\
  \bibnamefont {Rachel}}, \bibinfo {author} {\bibfnamefont {Christian}\
  \bibnamefont {Flindt}}, \bibinfo {author} {\bibfnamefont {Israel}\
  \bibnamefont {Klich}}, \bibinfo {author} {\bibfnamefont {Nicolas}\
  \bibnamefont {Laflorencie}}, \ and\ \bibinfo {author} {\bibfnamefont {Karyn}\
  \bibnamefont {Le~Hur}},\ }\bibfield  {title} {\enquote {\bibinfo {title}
  {Bipartite fluctuations as a probe of many-body entanglement},}\ }\href
  {\doibase 10.1103/PhysRevB.85.035409} {\bibfield  {journal} {\bibinfo
  {journal} {Phys. Rev. B}\ }\textbf {\bibinfo {volume} {85}},\ \bibinfo
  {pages} {035409} (\bibinfo {year} {2012})}\BibitemShut {NoStop}%
\bibitem [{\citenamefont {P\"oyh\"onen}\ \emph {et~al.}(2022)\citenamefont
  {P\"oyh\"onen}, \citenamefont {Moghaddam},\ and\ \citenamefont
  {Ojanen}}]{poyhonen2021observing}%
  \BibitemOpen
  \bibfield  {author} {\bibinfo {author} {\bibfnamefont {Kim}\ \bibnamefont
  {P\"oyh\"onen}}, \bibinfo {author} {\bibfnamefont {Ali~G.}\ \bibnamefont
  {Moghaddam}}, \ and\ \bibinfo {author} {\bibfnamefont {Teemu}\ \bibnamefont
  {Ojanen}},\ }\bibfield  {title} {\enquote {\bibinfo {title} {Many-body
  entanglement and topology from uncertainties and measurement-induced
  modes},}\ }\href {\doibase 10.1103/PhysRevResearch.4.023200} {\bibfield
  {journal} {\bibinfo  {journal} {Phys. Rev. Research}\ }\textbf {\bibinfo
  {volume} {4}},\ \bibinfo {pages} {023200} (\bibinfo {year}
  {2022})}\BibitemShut {NoStop}%
\bibitem [{\citenamefont {Bernien}\ \emph {et~al.}(2017)\citenamefont
  {Bernien}, \citenamefont {Schwartz}, \citenamefont {Keesling}, \citenamefont
  {Levine}, \citenamefont {Omran}, \citenamefont {Pichler}, \citenamefont
  {Choi}, \citenamefont {Zibrov}, \citenamefont {Endres}, \citenamefont
  {Greiner} \emph {et~al.}}]{51_atom_simulator}%
  \BibitemOpen
  \bibfield  {author} {\bibinfo {author} {\bibfnamefont {Hannes}\ \bibnamefont
  {Bernien}}, \bibinfo {author} {\bibfnamefont {Sylvain}\ \bibnamefont
  {Schwartz}}, \bibinfo {author} {\bibfnamefont {Alexander}\ \bibnamefont
  {Keesling}}, \bibinfo {author} {\bibfnamefont {Harry}\ \bibnamefont
  {Levine}}, \bibinfo {author} {\bibfnamefont {Ahmed}\ \bibnamefont {Omran}},
  \bibinfo {author} {\bibfnamefont {Hannes}\ \bibnamefont {Pichler}}, \bibinfo
  {author} {\bibfnamefont {Soonwon}\ \bibnamefont {Choi}}, \bibinfo {author}
  {\bibfnamefont {Alexander~S}\ \bibnamefont {Zibrov}}, \bibinfo {author}
  {\bibfnamefont {Manuel}\ \bibnamefont {Endres}}, \bibinfo {author}
  {\bibfnamefont {Markus}\ \bibnamefont {Greiner}},  \emph {et~al.},\
  }\bibfield  {title} {\enquote {\bibinfo {title} {Probing many-body dynamics
  on a 51-atom quantum simulator},}\ }\href {\doibase 10.1038/nature24622}
  {\bibfield  {journal} {\bibinfo  {journal} {Nature}\ }\textbf {\bibinfo
  {volume} {551}},\ \bibinfo {pages} {579--584} (\bibinfo {year}
  {2017})}\BibitemShut {NoStop}%
\bibitem [{\citenamefont {Zhang}\ \emph {et~al.}(2017)\citenamefont {Zhang},
  \citenamefont {Pagano}, \citenamefont {Hess}, \citenamefont {Kyprianidis},
  \citenamefont {Becker}, \citenamefont {Kaplan}, \citenamefont {Gorshkov},
  \citenamefont {Gong},\ and\ \citenamefont {Monroe}}]{53_atom_simulator}%
  \BibitemOpen
  \bibfield  {author} {\bibinfo {author} {\bibfnamefont {Jiehang}\ \bibnamefont
  {Zhang}}, \bibinfo {author} {\bibfnamefont {Guido}\ \bibnamefont {Pagano}},
  \bibinfo {author} {\bibfnamefont {Paul~W}\ \bibnamefont {Hess}}, \bibinfo
  {author} {\bibfnamefont {Antonis}\ \bibnamefont {Kyprianidis}}, \bibinfo
  {author} {\bibfnamefont {Patrick}\ \bibnamefont {Becker}}, \bibinfo {author}
  {\bibfnamefont {Harvey}\ \bibnamefont {Kaplan}}, \bibinfo {author}
  {\bibfnamefont {Alexey~V}\ \bibnamefont {Gorshkov}}, \bibinfo {author}
  {\bibfnamefont {Z-X}\ \bibnamefont {Gong}}, \ and\ \bibinfo {author}
  {\bibfnamefont {Christopher}\ \bibnamefont {Monroe}},\ }\bibfield  {title}
  {\enquote {\bibinfo {title} {Observation of a many-body dynamical phase
  transition with a 53-qubit quantum simulator},}\ }\href {\doibase
  10.1038/nature24654} {\bibfield  {journal} {\bibinfo  {journal} {Nature}\
  }\textbf {\bibinfo {volume} {551}},\ \bibinfo {pages} {601--604} (\bibinfo
  {year} {2017})}\BibitemShut {NoStop}%
\bibitem [{\citenamefont {Ebadi}\ \emph {et~al.}(2021)\citenamefont {Ebadi},
  \citenamefont {Wang}, \citenamefont {Levine}, \citenamefont {Keesling},
  \citenamefont {Semeghini}, \citenamefont {Omran}, \citenamefont {Bluvstein},
  \citenamefont {Samajdar}, \citenamefont {Pichler}, \citenamefont {Ho} \emph
  {et~al.}}]{256_atom_simulator}%
  \BibitemOpen
  \bibfield  {author} {\bibinfo {author} {\bibfnamefont {Sepehr}\ \bibnamefont
  {Ebadi}}, \bibinfo {author} {\bibfnamefont {Tout~T}\ \bibnamefont {Wang}},
  \bibinfo {author} {\bibfnamefont {Harry}\ \bibnamefont {Levine}}, \bibinfo
  {author} {\bibfnamefont {Alexander}\ \bibnamefont {Keesling}}, \bibinfo
  {author} {\bibfnamefont {Giulia}\ \bibnamefont {Semeghini}}, \bibinfo
  {author} {\bibfnamefont {Ahmed}\ \bibnamefont {Omran}}, \bibinfo {author}
  {\bibfnamefont {Dolev}\ \bibnamefont {Bluvstein}}, \bibinfo {author}
  {\bibfnamefont {Rhine}\ \bibnamefont {Samajdar}}, \bibinfo {author}
  {\bibfnamefont {Hannes}\ \bibnamefont {Pichler}}, \bibinfo {author}
  {\bibfnamefont {Wen~Wei}\ \bibnamefont {Ho}},  \emph {et~al.},\ }\bibfield
  {title} {\enquote {\bibinfo {title} {Quantum phases of matter on a 256-atom
  programmable quantum simulator},}\ }\href@noop {} {\bibfield  {journal}
  {\bibinfo  {journal} {Nature}\ }\textbf {\bibinfo {volume} {595}},\ \bibinfo
  {pages} {227--232} (\bibinfo {year} {2021})}\BibitemShut {NoStop}%
\bibitem [{\citenamefont {Steinhauer}(2016)}]{steinhauer2016observation}%
  \BibitemOpen
  \bibfield  {author} {\bibinfo {author} {\bibfnamefont {Jeff}\ \bibnamefont
  {Steinhauer}},\ }\bibfield  {title} {\enquote {\bibinfo {title} {Observation
  of quantum hawking radiation and its entanglement in an analogue black
  hole},}\ }\href {\doibase 10.1038/nphys3863} {\bibfield  {journal} {\bibinfo
  {journal} {Nature Physics}\ }\textbf {\bibinfo {volume} {12}},\ \bibinfo
  {pages} {959--965} (\bibinfo {year} {2016})}\BibitemShut {NoStop}%
\bibitem [{\citenamefont {Mu{\~n}oz~de Nova}\ \emph {et~al.}(2019)\citenamefont
  {Mu{\~n}oz~de Nova}, \citenamefont {Golubkov}, \citenamefont {Kolobov},\ and\
  \citenamefont {Steinhauer}}]{munoz2019observation}%
  \BibitemOpen
  \bibfield  {author} {\bibinfo {author} {\bibfnamefont {Juan~Ram{\'o}n}\
  \bibnamefont {Mu{\~n}oz~de Nova}}, \bibinfo {author} {\bibfnamefont
  {Katrine}\ \bibnamefont {Golubkov}}, \bibinfo {author} {\bibfnamefont
  {Victor~I}\ \bibnamefont {Kolobov}}, \ and\ \bibinfo {author} {\bibfnamefont
  {Jeff}\ \bibnamefont {Steinhauer}},\ }\bibfield  {title} {\enquote {\bibinfo
  {title} {Observation of thermal hawking radiation and its temperature in an
  analogue black hole},}\ }\href {\doibase 10.1038/s41586-019-1241-0}
  {\bibfield  {journal} {\bibinfo  {journal} {Nature}\ }\textbf {\bibinfo
  {volume} {569}},\ \bibinfo {pages} {688--691} (\bibinfo {year}
  {2019})}\BibitemShut {NoStop}%
\bibitem [{\citenamefont {Hu}\ \emph {et~al.}(2019)\citenamefont {Hu},
  \citenamefont {Feng}, \citenamefont {Zhang},\ and\ \citenamefont
  {Chin}}]{hu2019Unruh}%
  \BibitemOpen
  \bibfield  {author} {\bibinfo {author} {\bibfnamefont {Jiazhong}\
  \bibnamefont {Hu}}, \bibinfo {author} {\bibfnamefont {Lei}\ \bibnamefont
  {Feng}}, \bibinfo {author} {\bibfnamefont {Zhendong}\ \bibnamefont {Zhang}},
  \ and\ \bibinfo {author} {\bibfnamefont {Cheng}\ \bibnamefont {Chin}},\
  }\bibfield  {title} {\enquote {\bibinfo {title} {Quantum simulation of unruh
  radiation},}\ }\href {\doibase 10.1038/s41567-019-0537-1} {\bibfield
  {journal} {\bibinfo  {journal} {Nature Physics}\ }\textbf {\bibinfo {volume}
  {15}},\ \bibinfo {pages} {785--789} (\bibinfo {year} {2019})}\BibitemShut
  {NoStop}%
\bibitem [{\citenamefont {Kokail}\ \emph {et~al.}(2021)\citenamefont {Kokail},
  \citenamefont {Sundar}, \citenamefont {Zache}, \citenamefont {Elben},
  \citenamefont {Vermersch}, \citenamefont {Dalmonte}, \citenamefont {van
  Bijnen},\ and\ \citenamefont {Zoller}}]{Kokail2022}%
  \BibitemOpen
  \bibfield  {author} {\bibinfo {author} {\bibfnamefont {Christian}\
  \bibnamefont {Kokail}}, \bibinfo {author} {\bibfnamefont {Bhuvanesh}\
  \bibnamefont {Sundar}}, \bibinfo {author} {\bibfnamefont {Torsten~V.}\
  \bibnamefont {Zache}}, \bibinfo {author} {\bibfnamefont {Andreas}\
  \bibnamefont {Elben}}, \bibinfo {author} {\bibfnamefont {Beno\^{\i}t}\
  \bibnamefont {Vermersch}}, \bibinfo {author} {\bibfnamefont {Marcello}\
  \bibnamefont {Dalmonte}}, \bibinfo {author} {\bibfnamefont {Rick}\
  \bibnamefont {van Bijnen}}, \ and\ \bibinfo {author} {\bibfnamefont {Peter}\
  \bibnamefont {Zoller}},\ }\bibfield  {title} {\enquote {\bibinfo {title}
  {Quantum variational learning of the entanglement hamiltonian},}\ }\href
  {\doibase 10.1103/PhysRevLett.127.170501} {\bibfield  {journal} {\bibinfo
  {journal} {Phys. Rev. Lett.}\ }\textbf {\bibinfo {volume} {127}},\ \bibinfo
  {pages} {170501} (\bibinfo {year} {2021})}\BibitemShut {NoStop}%
\end{thebibliography}%

\end{document}